\documentclass[useAMS,usenatbib]{mn2e}
\usepackage{amsmath}
\usepackage{hyperref}
\hypersetup{
    colorlinks=True,
    linkcolor=red,
    filecolor=magenta,
    urlcolor=cyan,
    citecolor=blue,
}
\usepackage{graphicx}
\usepackage{natbib}
\usepackage{times}
\usepackage{float}
\usepackage{subcaption}
\usepackage{multirow}
\usepackage{color,soul}
\usepackage{import}
\usepackage[T1]{fontenc}
\usepackage{ae,aecompl}
\usepackage{amssymb}	
\usepackage{multicol}        
\usepackage{pgffor}
\usepackage{commath}	

\newcommand{\boldit}[1]{\textbf{\textit{#1}}}

\begin{document}

\defcitealias{Hossein12}{T12}
\defcitealias{Kinney96}{K96}
\defcitealias{Noll09}{N09}

\title[Classifying high-$z$ galaxy spectra]{Classifying galaxy spectra at $0.5<z<1$ with self-organizing maps}

\date{\today}
\author[S.~Rahmani, H.~Teimoorinia and P.~Barmby]
{S.~Rahmani$^{1,2}$\thanks{E-mail: srahma49@uwo.ca},
H.~Teimoorinia$^{3}$,
P. Barmby$^{1,2}$\\
$^{1}$Department of Physics $\&$ Astronomy, Western University, London, ON N6A 3K7, Canada\\
$^{2}$Centre for Planetary Science \& Exploration, Western University, London, ON N6A 3K7, Canada\\
$^{3}$NRC Herzberg Astronomy and Astrophysics, 5071 West Saanich Road, Victoria, BC, V9E 2E7, Canada
}
\maketitle

\begin{abstract}
    The spectrum of a galaxy contains information about its physical properties.
    Classifying spectra using templates helps elucidate the nature of a galaxy's energy sources.
    In this paper, we investigate the use of self-organizing maps in classifying galaxy spectra against templates.
    We trained semi-supervised self-organizing map networks using a set of
    templates covering the wavelength range from far ultraviolet to near infrared.
    The trained networks were used to classify the spectra of a sample of 142 galaxies with $0.5 < z < 1$ and the results compared to classifications performed using K-means clustering, a supervised neural network, and chi-squared minimization.
    Spectra corresponding to quiescent galaxies were more likely to be classified similarly by all methods while starburst spectra showed more variability.
    Compared to classification using chi-squared minimization or the supervised neural network, the galaxies classed together by the self-organizing map had more similar spectra.
     The class ordering provided by the one-dimensional self-organizing maps corresponds to an ordering in physical properties, a potentially important feature for the exploration of large datasets.
\end{abstract}

\section{Introduction}
\label{sec: intro_somz}

Nearly all information we can obtain from a galaxy is encapsulated in the light it emits; every observable phenomenon in a galaxy leaves a footprint on its spectral energy distribution (SED).
We can determine some physical properties of a galaxy by properly modeling various features observed in its SED, and the general shape of the SED can be used as an identifier of the morphological type.
Considering the main features of SEDs, galaxies can be categorized into two main groups: quiescent or star-forming.
Each group has its own characteristic features and can in turn be divided into many sub-branches.

Several attempts have been made to create a complete set of galaxy spectral templates using observations of nearby galaxies~(e.g.~\citealt{Kinney93}, \citealt[][hereafter \citetalias{Kinney96}]{Kinney96}, \citealt{Bershady00}, \citealt{Mannucci01}).
Based on their usage, these templates are restricted to certain wavelengths.
In the near-infrared (NIR) to ultraviolet (UV) wavelengths (the region where the energy output of stars peaks), the spectrum%
\footnote{In the remainder of this paper, we are concerned with a limited range in wavelength, and so we refer to spectra rather than spectral energy distributions.}
contains information about the main physical properties of galaxies, e.g., age, star formation rate (SFR), stellar mass, metallicity, and interstellar medium.

One approach to categorizing observed galaxy spectra involves matching to templates.
This approach can help to answer the question of whether galaxy spectra form a continuous distribution or can be separated into discrete groups.
Templates are also helpful in computing $K$-corrections and discovering new classes of objects as outliers \citep{Folkes96}.
A commonly used method of categorization via template matching is chi-squared minimization: the quantity
$\chi^2 = \Sigma_{\lambda} [({\rm obs} - {\rm templ})/{\rm uncert}]^2$
is calculated between each observation and each template and the best-matched template is that with the lowest $\chi^2$.
(For spectra, the observations and templates are typically made to match at a set/range of particular wavelengths, since the spectral shape is usually more important than the normalization.)
This technique is often used in computing photometric redshifts for galaxies \citep[e.g.][]{Hutchinson2016,Delchambre2016,Ilbert2006}.
In this study of galaxy spectral classification, we use chi-squared matching as a reference, since it is both simple and familiar to astronomers.
However, chi-squared matching has its limitations: it assumes that the templates cover
the full range of galaxy properties, and does not account for the possibility that a spectrum may share features with more than one template.
Uncertainties which are non-Gaussian or not correctly estimated may cause chi-squared matching  to fail.

Other approaches to classifying galaxy spectra include
artificial neural networks, K-means clustering, and principal component analysis (PCA)  \citep[e.g.][]{Karampelas2012,Allen13,Ordov14,Shi15}.
These methods can make use of templates, but generally do so in a more flexible manner than chi-squared matching, for example by giving the probability that a galaxy belongs to each of several categories.
Compared to chi-squared matching, these methods can provide additional insights about  observed spectra,
for example about their distances and/or ordering in feature space (neural networks, K-means) or their underlying dimensionality (PCA).
However, they have the disadvantage that they do not usually account for measurement uncertainties.

Neural networks can be trained using two methods: supervised and unsupervised.
In supervised methods, a neural network is trained using input data based on a desired outcome.
These methods are very useful for classification of data with specific target values.
In unsupervised methods there is no prediction of output:
these methods classify data based on their underlying structures and hidden patterns.
Unsupervised methods are very helpful for data exploration and finding hidden patterns when the underlying structure of the data is not well established.
A Kohonen self-organizing map (also called self-organizing map, or SOM) is a (semi)-supervised neural network.

The goal of this work is to investigate the use of self-organizing maps in classifying galaxy spectra against templates, and compare them to other methods.
The questions we want to answer are:
(1) What is the degree of agreement between classifications produced by $\chi^2$ matching, one- and two-dimensional self-organizing maps, K-means clustering and artificial neural networks?
(2) Are certain spectral types more (or less) consistently classified by the different methods?
(3) Which classification methods produce classes with the highest internal similarity?
(4) Does the ordering of classifications provided by one-dimensional self-organizing maps correspond to an ordering in physical properties?

We analyze the same sample of 142 galaxies with $0.5 < z < 1$ as
\citet[][hereafter \citetalias{Hossein12}]{Hossein12}, who classified the spectra using a supervised (multi-layer feed-forward) neural network method with the \citetalias{Kinney96} templates.
As a base for comparing the methods' performance, we find the best chi-squared match among the 12 templates for each of the 142 galaxies.
We use this to compute a chi-squared agreement score for each method.
To measure how well each classification method groups similar objects together, we
calculate the silhouette scores and dispersions in properties between galaxies classified in the same groups.
An important limitation to this analysis is the fact that {\em a priori} classifications for the observed galaxy spectra are not available.
This means that many classification metrics, which rely on knowledge of ``true'' labels, are not applicable.

This paper is organized as follows.
In Section~\ref{sec: data_highZ}, we present the data used to train the networks and the data classified using the trained networks.
We describe the SOM and K-means clustering methods in Section~\ref{sec: method_somz}.
The results of the spectral classifications and comparison with previous studies are presented in Section~\ref{sec: result}.
In Section~\ref{sec: summary_SOMZ}, we summarize our results and discuss potential future work in this subject.

\section{DATA}
\label{sec: data_highZ}
We use spectral templates from \citetalias{Kinney96} (in 10~\AA\ wavelength bins) to train neural networks.
We classify the spectra of 142 galaxies at $0.5<z<1$ from \citetalias{Hossein12} using the trained networks, and use their physical properties to test the new classifications.
Following the \citetalias{Hossein12} work, we chose these two sets of data not only to show the application of SOMs in spectral clustering, but also to easily compare supervised and unsupervised methods.

 \subsection{Kinney spectral templates}
     \begin{figure}
        \centering
        \includegraphics[width=0.47\textwidth]{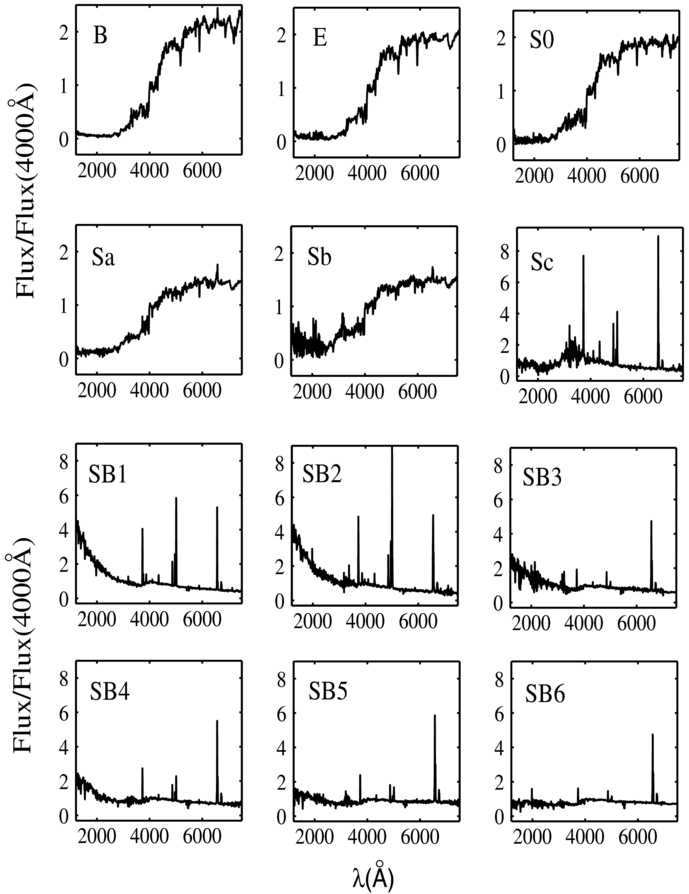}
        \caption {\citet{Kinney96} spectral templates for 12 types of galaxies, from fig. 1 in \citet{Hossein12}. The type of each template is shown in each frame. Plots B, E, S0, Sa, Sb and Sc show spectra that belong to quiescent galaxies. Starburst galaxy spectra are indicated with SB1 to SB6. Higher numbers represent more intrinsic extinction. Note that the vertical axis label from \citet{Hossein12} is incorrect:  the quantity plotted in the vertical axis is flux density, not flux.}
        \label{fig: k96}
    \end{figure}

    \citetalias{Kinney96} used ultraviolet-optical spectra of 70 star-forming and quiescent nearby galaxies to produce a set of templates that contained 12 types of spectral templates.
    These templates have been widely used in many studies to determine morphological type of galaxies or properties of specific types of galaxies~\citep[e.g.][]{Shakouri16, Paiano16, Laporte16, Holden16}.
    \citetalias{Kinney96} stated that these templates can also be used to classify the spectra of high-redshift galaxies.

    The 12 templates are divided based on their morphological types for quiescent galaxies or their extinction for starburst galaxies (Fig.~\ref{fig: k96}).
    The quiescent group of galaxies includes Bulge (B), Elliptical (E), S0, Sa, Sb, and Sc galaxies.
    The bulge group represents galaxies similar to M31 and M81, whose UV and optical spectra are dominated by their bulge stellar populations.
    The starburst galaxies are divided into six groups (SB1 to SB6) based on their intrinsic extinctions ($E(B-V)$).
    As Fig.~\ref{fig: k96} shows, SB1 galaxies have lower internal extinctions ($E(B-V) \simeq 0.05$), while SB6 galaxies have the highest amount of extinction ($E(B-V) \simeq 0.65$) among starburst galaxies.
    In the quiescent (B to Sb) templates, the spectrum is redder; strong absorption lines and the 4000~\AA~break are distinguishable.
    The SEDs of starburst galaxies are flatter in the optical and near-infrared region than those of the quiescent ones and show strong emission lines.
    For more details on each spectral type, we encourage readers to see \citetalias{Kinney96} and references therein.
   The \citetalias{Kinney96} spectra span from $\sim1200$~\AA~to $10000$~\AA~with a resolution of $\sim 10$~\AA.
    However, in this work we only use observations in the rest-frame wavelength range $\sim1200< \lambda < 8000$~\AA~to train our networks;
    this wavelength range was chosen due to the availability of flux information in those wavelengths for all 12 templates.

 \subsection{SED and Properties of the Sample Galaxies}
    \citetalias{Hossein12} selected 142 galaxies from the spectroscopic campaign of the ESO GOODS-South field~\citep{Vanzella05, Vanzella06, Vanzella08}.
    The 142 galaxies were selected based on the availability of photometry from HST/ACS, VLA/ISAAC, and {\it Spitzer}/MIPS and IRAC (10--13 filters with $\sim 0.4<\lambda<24~\mu$m in the observed frame).
   Data from these instruments was necessary in order to have a complete picture of stellar population and star formation rate.
    For each galaxy, a robust spectroscopic redshift and photometric measurements from the GOODS-MUSIC catalogue \citep{Santini09} were available.
\citetalias{Hossein12} used point-spread-function matched photometry
    from \citet{Santini09} as input to the Code Investigating GALaxy Emission ({\em CIGALE});~\citep[][hereafter N09]{Noll09} to generate the best-fit SED for each galaxy.
    \citetalias{Hossein12} produced the best SED match, with a wavelength interval of 910~\AA~to $\sim 80$~cm, for each galaxy\footnote{This wavelength interval is the default output of the {\em CIGALE} code.}.
    Assuming decreasing SFR and visual attenuation ($\tau$) model, Salpeter initial mass function~\citep{Salpeter55}, and an old stellar population with age of $\sim 10$~Gyr, he derived physical properties of the galaxies such as age and stellar mass.
    Some of these properties are shown in Table~\ref{tab: props}.
    In Section~\ref{sec: 1D_somz}, we study these properties for each category.
    More details on creating SEDs and extracting information about galaxy properties using {\em CIGALE} can be found in \citetalias{Noll09} and \citetalias{Hossein12}.

\begin{table}
\caption[Description of the properties of \citet{Hossein12} galaxies]{Description of the properties of \citet{Hossein12} galaxies; the output result of {\em CIGALE}}
\label{tab: props}
\centering
\begin{tabular}{l l l}
\hline\hline
\noalign{\smallskip}
Par. & Unit & Description\\
\noalign{\smallskip}
\hline
\noalign{\smallskip}
$f_\mathrm{burst}$ & --- & mass fraction of \\
& & young single population (SP) model \\
\noalign{\smallskip}
$t_{\,\mathrm{oSP}}$ & Gyr & age of old SP model \\
$t_{\,\mathrm{ySP}}$ & Gyr & age of young SP model \\
$t_{\,\mathrm{D4000}}$ & Gyr & D4000-related age \\
\noalign{\smallskip}
$M_\mathrm{star}$ & M$_\odot$ & total stellar mass  \\
SFR & M$_\odot$/yr & instantaneous SFR  \\
$A_\mathrm{FUV}$ & mag & attenuation at 1500\,\AA{} \\
\noalign{\smallskip}
\hline
\end{tabular}
\end{table}

The basis for our classification is the SEDs that were produced by \citetalias{Hossein12}.
Both the raw data used to create the SEDs,%
\footnote{See this \href{http://telbib.eso.org/detail.php?bibcode=2012AJ....144..172T}{ESO webpage}.}
and the SEDs themselves, in the form of flux per rest frame wavelength, are publicly available
(see Appendix~\ref{app:sed_props} and the Supplementary Information).
Since we have chosen the \citetalias{Hossein12} SEDs to be classified by the trained network, we only used the part of the SEDs that have the same wavelength range as the \citetalias{Kinney96} spectral templates.

\section{METHOD}
\label{sec: method_somz}

In this section we  briefly review the use of
K-means clustering, artificial neural networks, and self-organizing maps in astronomy and describe their algorithms.
As K-means is both simpler and more well-known, we introduce it briefly first before discussing artificial neural networks in general and the special case of self-organizing maps in detail.

\subsection{K-means clustering}
\label{sec: kmeans_method}

K-means clustering~\citep{Macqueen67} is an unsupervised method used in many astronomical studies~\citep[e.g.][]{DAbrusco12,Ordov14,Boersma14,Aycha16}.
The K-means method partitions a data set $x_i$, having $n$ points in $d$-dimensional feature space, into $K$ clusters, in such a way that each data point belongs to the cluster with the nearest mean value.
The number of clusters $K\leq n$ must be decided by the user in advance, and the cluster centroids in feature space are located at positions $\mu_k$.
The goal of the algorithm is to minimize the sum
\begin{equation}
\label{eq:km_sos}
J = \sum_{i=1}^n \sum_{k=1}^K {\min}_k \big( \norm{x_i - \mu_k}^2 \big)
\end{equation}
where $\norm{z}$ indicates the distance measurement in $d$-dimensional space. In this work the Euclidean distance is is used, but other metrics are also possible.
The algorithm is initialized by randomly choosing $K$ points as cluster centroids.
Each data point is assigned to the cluster with the closest centroid, and the centroids are re-calculated.
The steps of re-assigning cluster membership and re-calculating the cluster centroids are repeated until a stopping criterion is reached (a maximum number of iterations or no change in cluster assignments).
We used the \textsc{matlab} K-means library~\citep{Seber84, Spath85}, written for \textsc{matlab2015b}, to perform  K-means clustering.
The default maximum number of iterations (100) was used.

\subsection{Artificial neural networks}
 \label{sec:ann}

Artificial neural networks \citep[ANNs;][]{bishop06}, which are inspired by the way neurons in a human brain route and process data, are very powerful tools that are used in data processing and pattern recognition problems.
An ANN contains many interconnected units (nodes or neurons) which process data and work together to solve problems.
It uses a training method to learn about nonlinear and complex relations between input and output data, and how to apply these relations to new sets of data.

One example of such a training method is back propagation, an error-correction learning method
in which the corrections are made in the last layer and propagate toward the first layer.
The signal for the corrections comes from a loss function (e.g., a mean square error function).
ANNs have many uses in statistics including as an alternative to conventional regression models;
\citet{Marquez91} found that ANNs ``perform best [in terms of mean absolute percentage error] under conditions of high noise and low sample size," conditions which often occur in astronomy.

ANNs have seen a number of uses in extragalactic astronomy.
These include object detection and star/galaxy classification \citep{Andreon00},
estimation of photometric redshifts \citep{vanzella04},
predicting morphological classifications, spectral types and redshifts for galaxies in the Sloan Digital Sky Survey \citep{Ball04},
ranking quenching parameters in central galaxies \citep{Hossein16a},
and predicting galaxy infrared luminosities from optical observations \citep{Ellison16a}.
Comparing an ANN-based technique to $\chi^2$ minimization for estimating reddening of O and B stars,
\citet{Gulati97} found that the two methods showed good agreement, with the ANN method showing faster performance on a test set.
\citet{vanzella04} found that, compared to other photometric redshift estimation techniques, their ANN-based technique performed better in terms of both redshift precision and computational speed.
In this work we are comparing to the results of \citetalias{Hossein12} who used a two-layer feed-forward ANN with a mean-squared error performance function, minimized via the Levenberg-Marquardt algorithm.
For further details of the ANN classification we refer the reader to \citetalias{Hossein12}.

 \subsection{Self-organizing maps}
 \label{sec: som}

A Kohonen self-organizing map (also called self-organizing map, or SOM) is a (semi)-supervised neural network for mapping and visualizing a complex and nonlinear high dimension data introduced by~\citet{Kohonen82}.
It shows simple geometrical relationships in non-linear high dimensional data on a map \citep{Kohonen98}.
The training of a self-organizing map is fully unsupervised;
a competitive learning method is used in which the neighborhood function serves to find the topological behaviour of the input data.
     Using the resulting map as a template to classify other data requires labelling the output, which is why some groups consider self-organizing maps to be a (semi)-supervised method.
The utilization of the SOM in astronomy dates back to the 1990s, with \citet[][]{Odewahn92}, \citet[][]{Hernandez94}, and \citet[][]{Murtagh95} among the first to use SOMs in their studies.
\citet{Geach12} used COSMOS data to demonstrate two of the main applications of SOMs: object classification and clustering, and photometric redshift estimation. The latter has been the subject of many other studies \citep[e.g.][]{Kind14a}.
From classifying quasars' spectra to star/galaxy classifications, from gamma-ray burst clustering to classification of light curves, this method has proved to be useful in various fields of astronomy \citep[e.g.][]{Maehoenen95, Miller96, Andreon00, Balastegui01, Rajaniemi02, Brett04, Scaringi09}.

Large spectroscopic surveys have made available integrated spectra of millions of galaxies.
These integrated spectra combine the light of billions of individual stars and nebulae within a galaxy, and
finding patterns and common characteristics between galaxies can be a complex task.
\citet{In12} introduced a new clustering tool based on the SOM method for analyzing these large datasets.
They used $\sim 60000$ spectra from the Sloan Digital Sky Survey \citep[SDSS;][]{Abazajian09} to test their tool, and created very large SOMs to analyze the type of spectra/objects.
They also generated SOMs from quasars' spectra in order to find unusual types of spectra.
Later, \citet{Meusinger16} used these SOMs and updated data from SDSS and other surveys to find a new class of quasars.
The other application of SOMs is to find outliers or errors in the data.
\citet{Fustes13} produced a package based on SOM to classify spectra from the GAIA survey that were previously classified as ``unknown'' by the SDSS pipeline. This package can distinguish an astronomical object from instrumental artifact, and then classify the object based on its spectrum.

The self-organizing map is a clustering method which reduces the dimensionality of the data, usually to one or two dimensions (1D or 2D), while preserving topological features of the original dataset.
 The result of an SOM is a set of nodes (neurons) that are arranged in a 1D or 2D arrays \citep{Kohonen98}. To first order, SOM and K-means work similarly: both algorithms operate on the distances between
$n$ objects, each represented by $d$ features.
The SOM algorithm also considers distances between objects in the low-dimensional map, as controlled by a neighbourhood function.
When the radius of the neighbourhood function goes to zero, the SOM algorithm loses its ordering power and acts like K-means clustering.
 Each node may contain one or more samples from the input data.
 The distance between the nodes represents similarity or dissimilarity of the underlying samples, i.e., similar data are closer together in the array and the distance between two nodes is related to the dissimilarity of their samples.
 A weight vector ``\boldit{W}" with the same dimension as the input data is associated with each node and will be varied during the training process.
 This vector is the key factor in determining the position of the nodes in a map.
The set of weight vectors  \boldit{W} $\in \Re^{md}$ where $m$ is the number of nodes.
 \cite{Geach12} presented the application of the self-organized map and discussed its algorithm in detail.

 \subsubsection{SOM algorithm}
 \label{sec: algorithm}
     Assume we have a dataset which contains vectors, \boldit{V} $\in \Re^{nd}$, and we want to map them on an S1 by S2 map.
     Sizes of SOMs are arbitrary and there are no rules regarding choosing one over the other.
    \citet{Vesanto05} suggested that a total number of $5\sqrt{d}$ neurons is a sufficient size, but users usually choose the size of the grids based on their dataset and their application of the results.

     We start by creating ${\rm S1} \times {\rm S2}=m$ empty neurons.
     The arrangement of these neurons depends on the map's topology provided by the user.
     In the case of 1D maps, since each neuron has two immediate neighbours, the topology of the map does not have any effect on the final result and any topology can be chosen.
     However, in 2D maps, the shape of the neurons specifies the number of immediate neighbours for each neuron and it is up to user to choose the most suitable shape based on the data.
     In this paper, we choose hexagonal topology, which gives each neuron six neighbours, and provides more interactions between neurons.
     Initially a random weight vector,  $ w \in \boldit{W}$, will be assigned to each node.
     The process of creating SOM happens over a series of $N$ iterations (indexed by timestep $t$), where $N$ is set by the user.
     During each iteration the weight vectors might change according to the Kohonen learning rule (equation~\ref{equ: weight adj}).
      In each iteration, the SOM code:
     \begin{enumerate}
        \item chooses a random vector from the dataset ($v \in V$).
        \item calculates the Euclidean distance in $\Re^d$ space for each node, $j$, as  $D_j^2= \sum_{i=0}^{i=d} (v_i - w_{ji})^2$, and finds a neuron with minimum $D_j$, (``$D_{j_{\min}}$"). This neuron is the winning node and is called the Best Matching Unit (BMU).
        \item  computes the radius of the neighbourhood of the BMU to find nodes within this radius. The weight vectors of these nodes will be affected in the next steps. The radius of the neighbourhood is arbitrary and can be set to be as high as half of the SOM size. It then decays exponentially over each iteration as
        \begin{equation}
            r^t_{\rm BMU} = r^0_{\rm BMU}e^{(-t/\tau)}
        \end{equation}
        where $\tau$ is a decay constant and is usually set to be the same as the number of iterations, $N$. $r^0_{BMU}$ and $r^t_{\rm BMU}$ are the radii of the neighbourhood at 0th and $t$th iteration, respectively.
        \item changes the weight vectors of the BMU and all the nodes within $r^t_{\rm BMU}$ as:
        \begin{equation}
            \label{equ: weight adj}
            w(t+1)=w(t)+L(t) \times R(t) \times(v(t)-w(t))
        \end{equation}
        where $L(t) = L_0 e^{(-t/\tau)}$ is the learning factor, which prevents the divergence of the SOM and $R(t)=\exp(-\frac{D_j^2}{2r^t_{\rm BMU}})$ is the influence rate. $R(t)$ determines how the weight of each node in the neighbourhood of BMU will change.
     \end{enumerate}
     These steps are then repeated $N$ times.

 \subsubsection{Creating self-organizing maps}
\label{sec: create_som}
     In order to create SOMs, we use the {\sc matlab} neural network toolbox~\citep[NNT,][]{matlabtolbox}, written for \textsc{matlab2015b}.
     An SOM in {\sc nnt} can be created by the {\sc newsom} or {\sc selforgmap} libraries, both of which work in two phases, an ``ordering phase" and a ``tuning phase".
     The first phase is called the ``ordering phase" and
     starts with maximum neighbourhood distance and an initial high learning factor (usually 0.9) is provided by the user.
     The ordering phase continues for a requested number of iterations.
     During the iterations, the learning factor decreases to the tuning phase learning factor and the neighbourhood distance reaches that of the tuning phase as well.
     Both the learning factor and the neighbourhood of the tuning phase are set by the user.
     The amount by which these two factors change in each iteration depends on the number of iterations.

     In the second, or ``tuning'' phase,
     the neighbourhood distance is kept at the user-defined minimum.
     The learning factor, however, decreases gradually.
     The gradual change in the leaning factor helps to fine-tune the topology results, leading to a more stable SOM.
     To allow the fine tuning, the number of iterations in this phase must be much larger than the that of the ordering phase.
     We chose the number of epochs in the tuning phase to be 3 times the number of epochs in the ordering phase.

     We create the final SOMs with initial values for number of iterations in ordering phase, ordering phase learning factor, tuning phase learning factor, and tuning phase neighbourhood distance of 1000, 0.9, 0.02, and 1, respectively.
     To present our results, we use {\sc nnt}'s built-in plotting tool.
     Specifically, we use two of the plots in this tool: a hits map, which shows the number of times each neuron has become the winner (hits), and a distance map, which shows the distance between those neurons.
     In the maps, the coloured hexagonal shapes represent the neurons.
     The distances in a distance map are shown by the grey cycle colours:
     the darker the colour, the larger the distance between neurons.
     In the hit maps, neurons with zero hits are left empty.

     As mentioned in the introduction, one of the main advantages of the SOM method is that the resulting networks can be used to cluster new datasets with no additional training required.
     New data with the same dimensionality as the original input data can be presented to the already trained network: the SOM algorithm finds the best matching unit considering the weight of nodes in the trained network and each vector from the new data set.
     The winning node determines the place of each new vector on the SOM.
     The location of the new vectors on the SOM allows them to be compared with the original data set.

     \subsection{Classification and clustering metrics}
     \label{sec:metrics}

As discussed in the Introduction, we compare the classification methods described above to the chi-squared template matching method.
We compute the value of chi-squared between each galaxy spectrum and each template spectrum as
\begin{equation}
\chi^2 = \Sigma_i ({\rm spectrum} - {\rm template})^2/{\rm template}
\end{equation}
where the $i$ are all of the tabulated wavelengths in spectra and templates.
This treatment is the equivalent of assuming that the \citetalias{Kinney96} template spectra are photon-noise dominated, since they do not have tabulated uncertainties.
Each galaxy spectrum is assigned to the template class with the lowest chi-squared value.

We use several metrics to measure the degree of agreement between classification methods. We compute a ``chi-squared agreement'' score as follows.
    If a method sorted a galaxy into the same cluster or neuron as its chi-square matched template, this was considered a match and scored 1 point.
    For the larger-sized self-organized maps, if a galaxy was classified between two nodes, one of which contained its best-fit chi-square template, this scored 0.5 point.
    Otherwise, the chi-squared agreement score for that galaxy was zero points.
Summing the chi-squared agreement score over all galaxies in the sample indicates how well each technique matched the results of chi-squared fitting.
Summing the chi-squared agreement score over all methods for a single galaxy indicates how well the methods agreed for a particular spectrum.

Classification and clustering are related techniques, and we make
another comparison between the classification methods with a clustering metric, the silhouette score  \citep{rousseeuw87}.
This score is a metric used to describe cluster compactness and isolation, given by:
\begin{equation}
S = \frac{1}{n} \sum{\frac{b - a}{\max\big(a, b\big)}}
\end{equation}
where $a$ is the mean distance between a point and the other points in its cluster, and $b$ is the mean distance between a point and the nearest cluster of which that point is not a member.
A higher silhouette score corresponds to a partition with better defined clusters.
For a given dataset, the score generally declines with the number of clusters and is sometimes used to determine the optimal number of clusters.
In this work we use the silhouette score to compare classifications with a comparable number of classes, as a measure of similarity within the classes.
Mean values for each classification were calculated using the \textsc{scikit-learn} Python package \citep{sklearn}.

\section{RESULTS AND DISCUSSION}
\label{sec: result}

    In this section we show the results of the neural networks and K-means clustering trained using the \citetalias{Kinney96} template spectra.
    Training with \citetalias{Kinney96} templates results in networks or clusters that have regions corresponding to galaxies of known morphological type.
    The trained networks or clusters can then be used to categorize other galaxies.

    The first step in categorizing galaxies is choosing the number of classes.
    In the K-means method empty clusters are not allowed;
    hence a $K=12$ clustering would produce one-object clusters as in Fig.~\ref{fig: k96}.
    We did not want to have too dense or too sparse clusters.
    Therefore we chose the number of clusters in the K-means clustering method to be $K=4$.
    Since empty neurons are allowed in SOMs, finding an optimal number of neurons for SOM networks is more subtle than just specifying a number of clusters.
    In order to find a sufficient size for the trained neural networks, we created maps with sizes ranging from $1\times2$ to $50\times50$.
    Varying the grid size of the maps helps us to monitor whether tighter grouping of galaxies is due to their similar properties or a lack of map space to separate them.
    Based on the size of the data and SOM results, we found the optimum grid size to be $1\times22$ and $12\times12$ in 1D and 2D maps, respectively.
    For each grid we created different SOMs with different learning factors, neighbourhood distances, and number of iterations to find the optimum.

    We started our analysis by creating 1D SOMs and K-means clustering.
    First, we created SOMs with only two neurons ($1\times2$ map), and then increased the number of neurons one at a time (Section~\ref{sec: 1Dt}).
    We performed K-means clustering on the \citetalias{Kinney96} templates (Section~\ref{sec: Kmeansvssom}) and compared
   the results with the results of the $1\times4$-sized 1D SOMs (Section~\ref{sec: Kmeansvssom}).
    For 2D networks (Section~\ref{sec: 2D}), we found the optimum size for the network to be $12\times12$ and created two types of SOM to show different applications of 2D SOMs.
    In Section~\ref{sec: Kmeansvssomvsann} we compare the results of the different classification in several ways.
    For each generated neural network, we compare the results with the \citetalias{Kinney96} categorization.
    We also use these networks to classify the \citetalias{Hossein12} galaxy sample, and compare this classification with that from the supervised networks in \citetalias{Hossein12} and with K-means clustering.

    \subsection{One-dimensional self-organizing maps}
    \label{sec: 1D_somz}
        \subsubsection{Training the networks}
        \label{sec: 1Dt}
            To start our classification, we assumed that galaxies can be divided into only two general types, quiescent and starburst.
            This corresponds to a network with only two neurons.
            We increased the size of the map gradually until the 12 input samples divide into the 12 neurons.

\begin{figure}
                \begin{subfigure}[b]{0.45\textwidth}
                    \centering
                  \includegraphics[width=\textwidth]{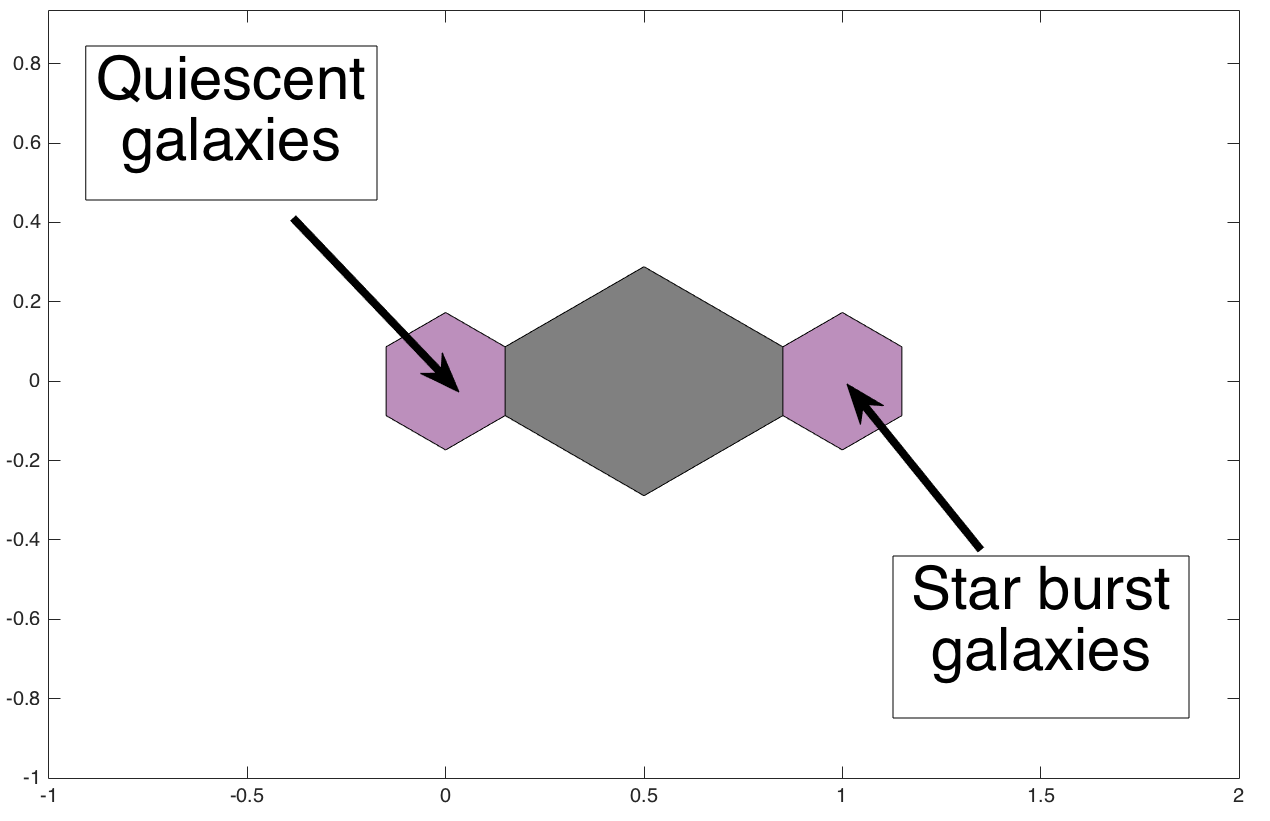}
                \end{subfigure}
                \hfill
                \begin{subfigure}[b]{0.45\textwidth}
                    \centering \includegraphics[width=\textwidth]{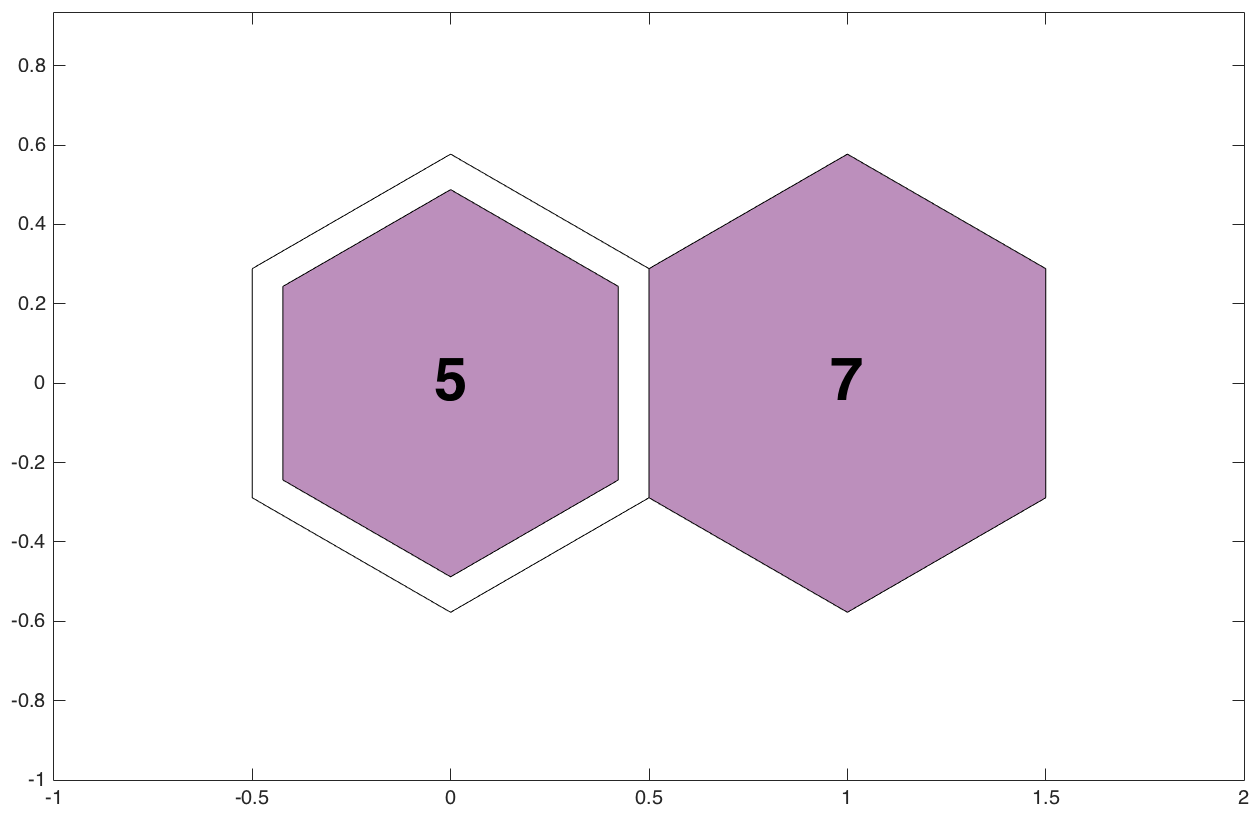}
                \end{subfigure}
                \caption[Results of training network in $1\times2$~grid]{Results of training network in $1\times2$~grid. The upper panel is a distance map and the lower panel is a hit map. In this network, 5 of the templates from \citet{Kinney96} are categorized as quiescent galaxies and the rest are starbursts. Because of their strong emission lines, Sc galaxies are moved towards the starburst ones.}
                 \label{fig: 1by2T}
            \end{figure}

\begin{figure}
                \begin{subfigure}[b]{0.45\textwidth}
                    \centering
                    \includegraphics[width=\textwidth]{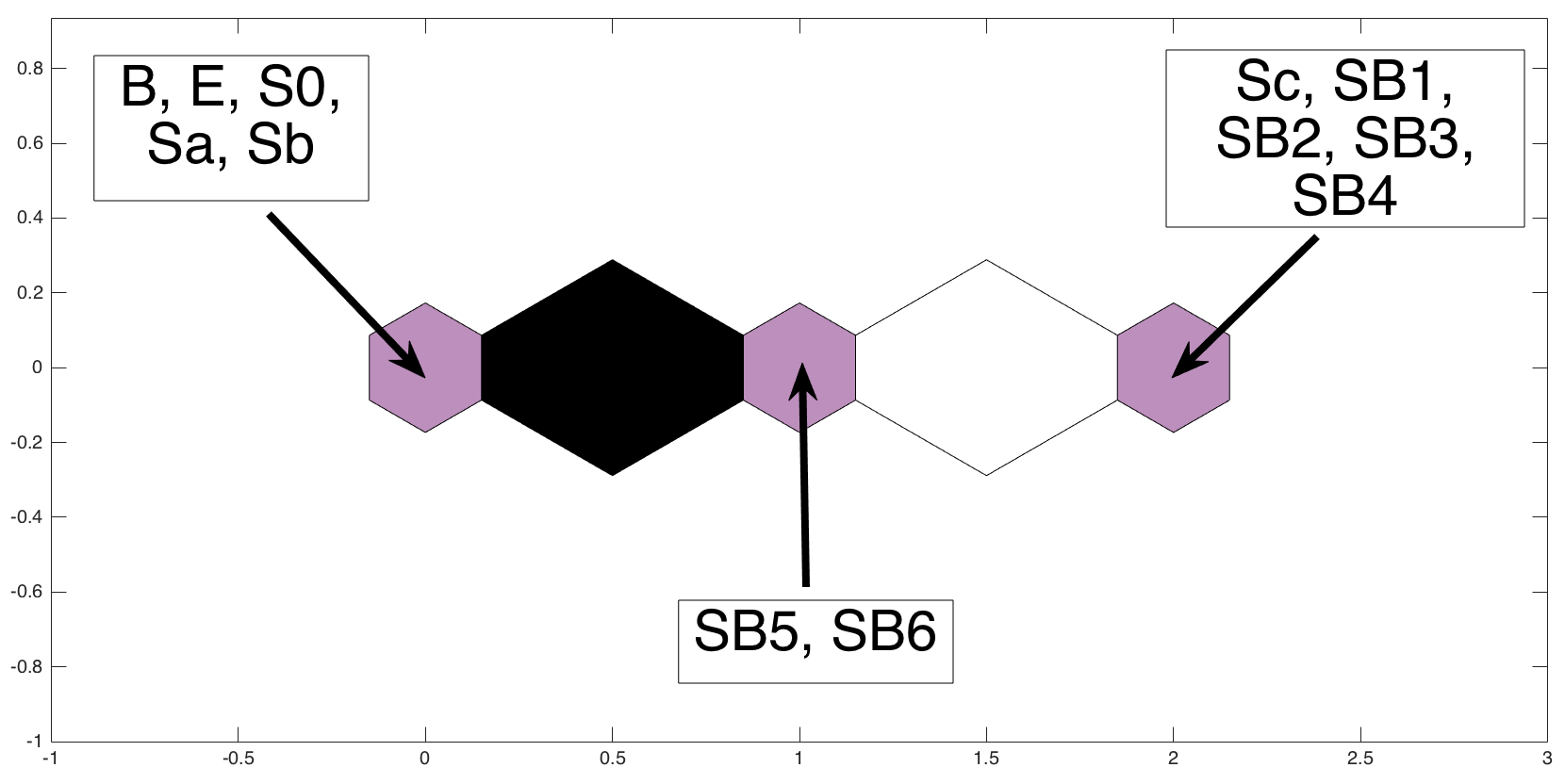}
                \end{subfigure}
                \hfill
                \begin{subfigure}[b]{0.45\textwidth}
                     \includegraphics[width=\textwidth]{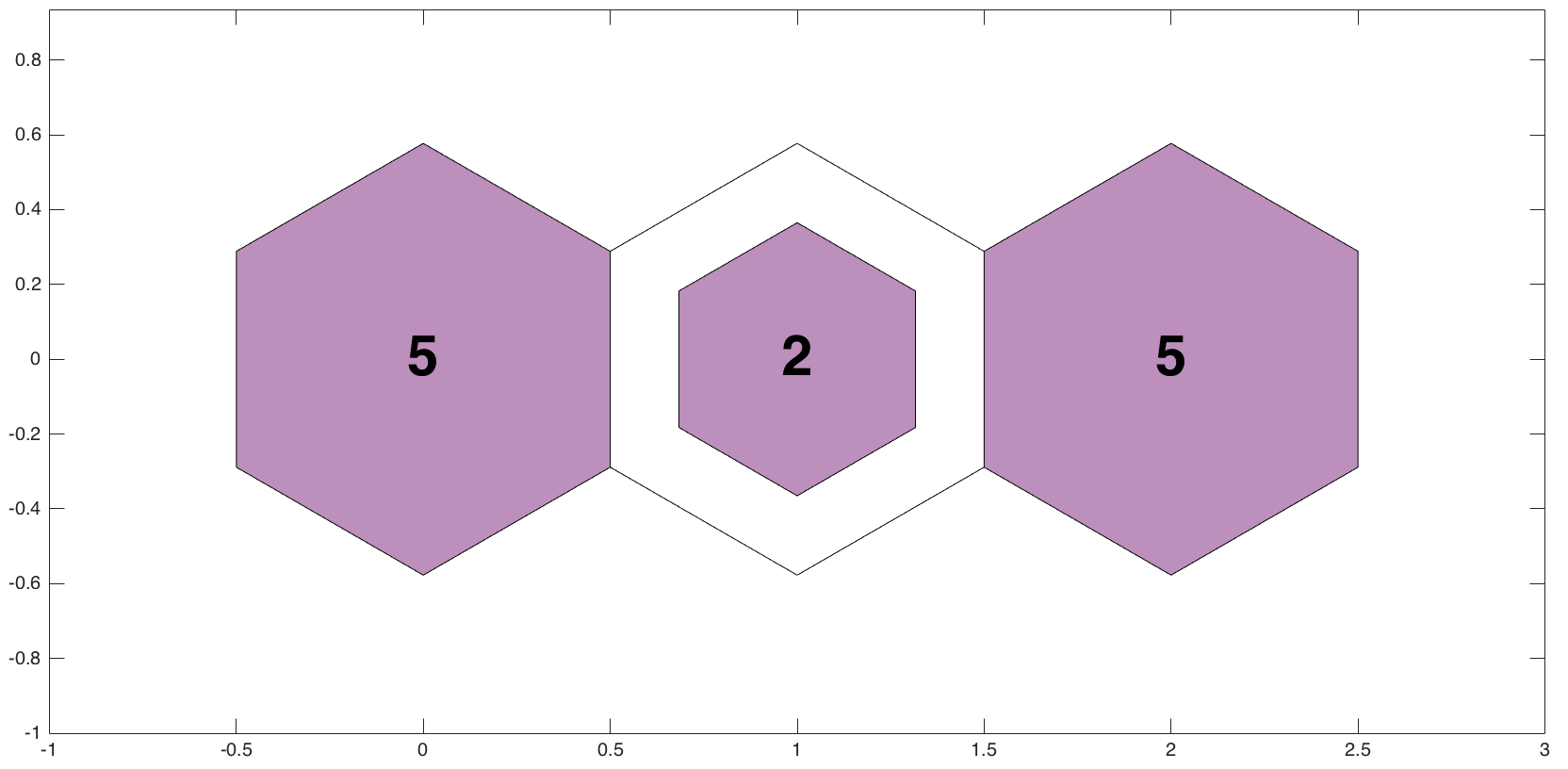}
                \end{subfigure}
                \caption[Results of training network in $1\times3$~grid]{The same as Fig.~\ref{fig: 1by2T} but showing the results of training network in $1\times3$~grid. In this network again, 5 of the \citet{Kinney96} templates are categorized as quiescent and 7 as starbursts. However, this time 2 templates (SB5 and SB6) are separated from the starburst groups.}
                 \label{fig: 1by3T}
            \end{figure}

\begin{figure*}
                \begin{subfigure}[b]{\textwidth}
                    \centering
                    \includegraphics[width=\textwidth]{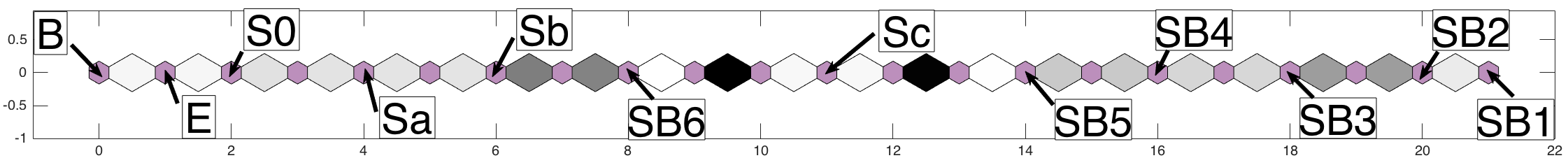}
                \end{subfigure}
                \hfill
                \begin{subfigure}[b]{\textwidth}
                    \includegraphics[width=\textwidth]{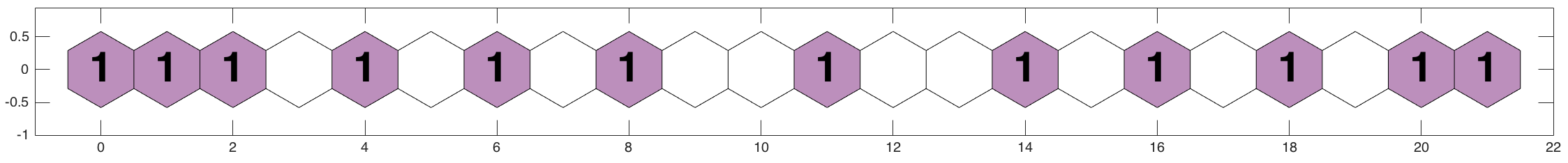}
                \end{subfigure}
                \caption[Results of training network in $1\times22$~grid]{The same as Fig.~\ref{fig: 1by2T} but this time the figure shows results of training network in $1\times22$~grid.}
                \label{fig: 1by22T}
            \end{figure*}

            Figs.~\ref{fig: 1by2T} --~\ref{fig: 1by22T} show the results of the training networks.
            The upper panels of the figures show the neurons and the relative distances between their weights.
            As mentioned in Section~\ref{sec: method_somz}, an increase in the darkness of colours between neurons represents an increase in relative distance between the neurons.
            The lower panels of the figures show the number of \citetalias{Kinney96} templates that are placed in each neuron.

            In the upper map in Fig.~\ref{fig: 1by2T}, the dark colour between two neurons indicates that the relative distance between these two groups is high, and these two groups are distinguishable.
            In the lower part of Fig.~\ref{fig: 1by2T}, we see that the templates are divided into two groups of 5 and 7.
            Although we know from \citetalias{Kinney96} and \citetalias{Hossein12} that 6 of the templates are quiescent and the other 6 are starbursts, the SOM results show 5 of the galaxies in one group and the other 7 in the second group.
            In this method, the Sc template has been categorized as starburst due to the relatively higher disk stellar population relative to that of the bulge.
            According to \citetalias{Kinney96}, Sc galaxies are considered to be late Hubble type galaxies which have  flatter spectra compared to other quiescent galaxies.

            Fig.~\ref{fig: 1by3T} shows the results of the training in a 1$\times$3 network.
            In these plots we force the galaxies to be categorized in a maximum of three groups.
            If the templates in Fig.~\ref{fig: 1by2T} truly belonged in two groups, they would be grouped into two groups in this network, even when we try to cluster them into three.
            However, in the lower panel of Fig.~\ref{fig: 1by3T}, we can see that the middle node contains two templates.
            These two templates, which are separated from the group of starburst templates in the lower part of Fig.~\ref{fig: 1by2T},  are the SB5 and SB6 types.
            In the upper plot of Fig.~\ref{fig: 1by3T}, the colour between two right neurons is black and the colour between two left neurons is white.
            The black colour indicates that the left neuron has the most differences with the other two groups,
            while the white colour shows that the two right neurons are very similar to one another.

            Comparing Fig.~\ref{fig: 1by2T} to Fig.~\ref{fig: 1by3T} shows that the starburst templates are divided into two groups.
            Based on the colours between these two groups, we conclude that they are more similar to each other than to the ones in the left neurons; both groups are starbursting and have strong emission lines.
            On the other hand, the SB5 and SB6 templates have the highest internal extinctions; this causes the spectra to become flatter at shorter wavelengths.
            The flatter UV spectra make these two templates more similar to quiescent galaxies than to other starbursts.
            Therefore, in the networks, SB5 and SB6 types are grouped close to the quiescent galaxy templates.

            We increased the size of the maps gradually until the templates were divided into twelve groups (Figs.~\ref{fig: 1by4T} to ~\ref{fig: 1by20T} in Appendix~\ref{app: high_Z_1d_soms} and Figure~\ref{fig: 1by22T}).
            Since there are more nodes in higher grid SOMs, the algorithm
            pays more attention to small differences between groups.
            If the templates from \citetalias{Kinney96} had completely distinct spectral types, then a $1\times12$-sized SOM would be expected to show 12 different groups each containing a single template.
            However, in Fig.~\ref{fig: 1by12T}, we see that three of the neurons contain two templates and three of them are empty.
            It is evident that there was no template in the \citetalias{Kinney96} sample that can fill those empty neurons.
            In Fig.~\ref{fig: 1by12T}, from left to right, templates with types B and E, SB3 and SB4, and SB1 and SB2 are the ones grouped together.
            The SB3 and SB4 grouping breaks when we increase the size of the network to $1\times15$~(Fig.~\ref{fig: 1by15T}).
            The SB1 and SB2 templates, however, remain in the same neuron until the size of the map is increased to $1\times20$~(Fig.~\ref{fig: 1by20T}).
            The separation between templates B and E only happens when the size of the SOM exceeds $1\times22$~(Fig.~\ref{fig: 1by22T}).

            In Fig.~\ref{fig: 1by22T}, we can see twelve distinct groups.
            Five groups in the left-side neurons are separated from the rest of the groups with two dark-grey colours between them.
            Five groups in the right-side neurons are separated from the rest by the black colour, and Sc and SB6 galaxies are in two separate blocks.
            This figure is a good example of the problem of using one-dimensional SOMs to explore the data.
            Since are the templates are forced to be arranged in a linear array, the SB6 and Sc groups are isolated and not connected to any other groups.
            However, in a two-dimensional map, due to the connection of each neuron to 6 more neurons, template galaxies have more degrees of freedom to find their relation with other neurons.
            As a big picture, Fig.~\ref{fig: 1by22T} shows that even though the galaxies are clustered in twelve groups, there still remain only two main distinct groups.
            The closest occupied neuron in the starburst side of the SOM, belongs to the SB6 type.
            This template has the most extinction and  the general shape of its spectrum has more similarity to quiescent galaxies than to other starburst types.

            The fact that the templates need to have at least 22 different neurons to be divided into 12 groups shows that the differences between SB1 and SB2, and B and E, templates are very small.
            Because of their similarities, they tend to stay in the same group until the network becomes large enough to pay attention to the smallest particularity.

          \subsubsection{One-dimensional self-organizing maps and K-means clustering: templates}
          \label{sec: Kmeansvssom}

    \begin{figure*}
    \begin{subfigure}[b]{0.49\textwidth}
        \centering
        \includegraphics[width=.99\textwidth, height=7.5cm]{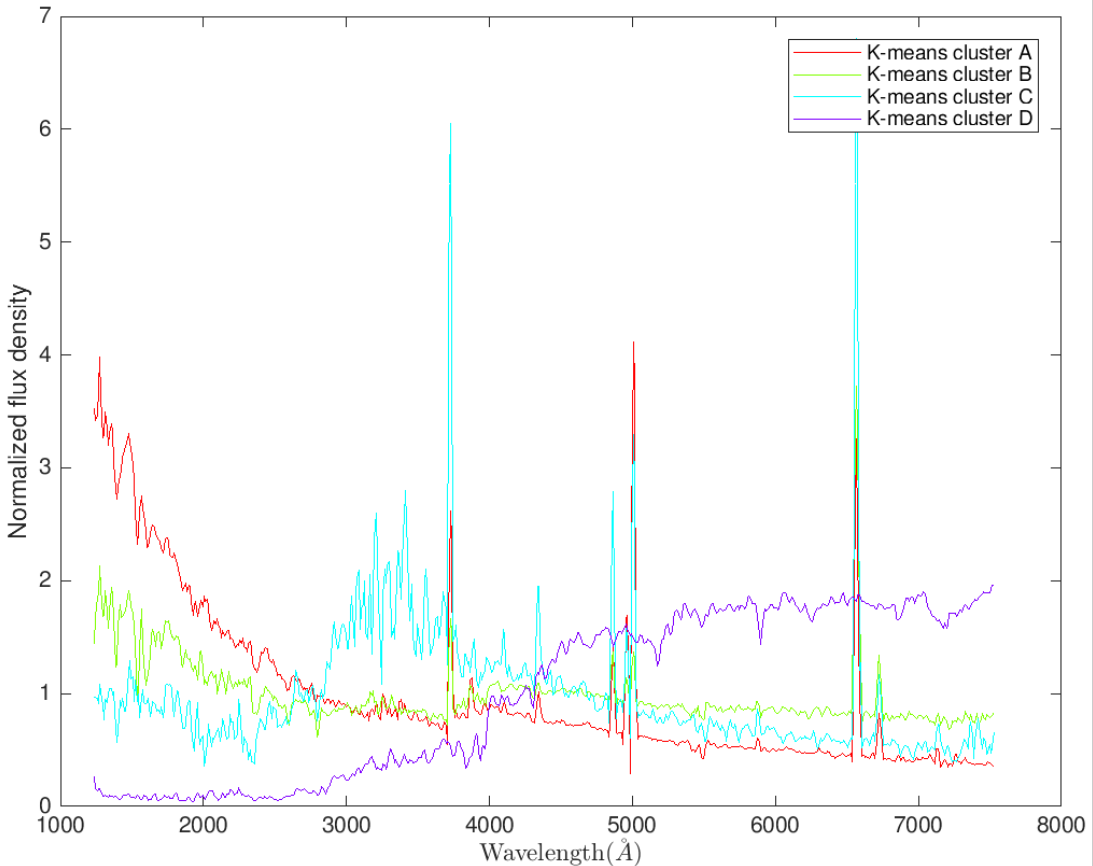}
    \end{subfigure}
    \hfill
        \begin{subfigure}[b]{0.49\textwidth}
        \centering \includegraphics[width=.99\textwidth, height=7.5cm]{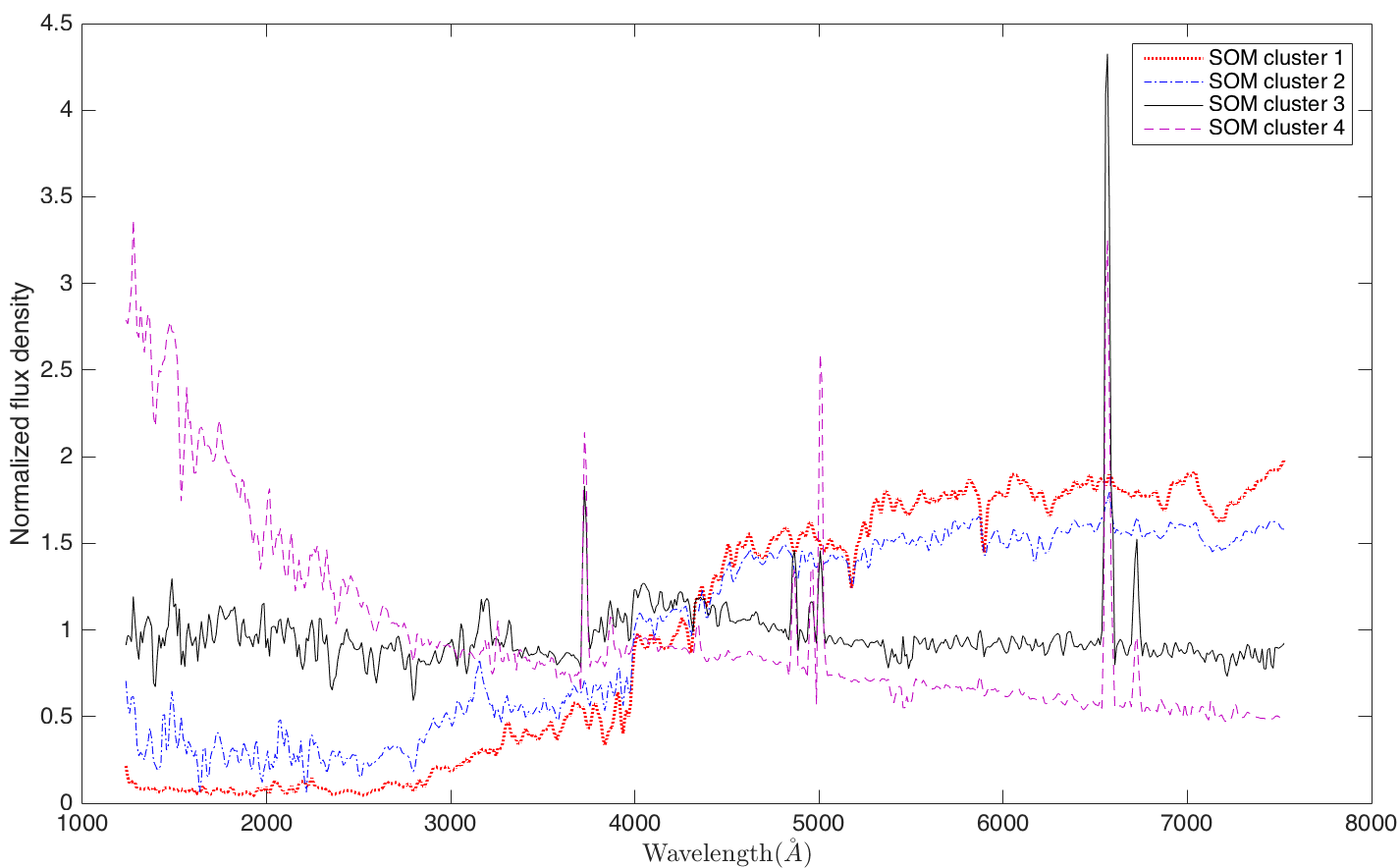}
    \end{subfigure}
    \caption{Clustering the \citetalias{Kinney96} template spectra using K-means clustering (left) and SOM (right). Note that both methods randomly assign the initial values for their analysis, therefore the cluster labels are only used for distinguishing cluster membership.}
    \label{fig: som_k_means_4}
    \end{figure*}

    We performed K-means clustering on the \citetalias{Kinney96} template spectra using 4 clusters.
    Fig.~\ref{fig: som_k_means_4} shows the clustering of the template spectra into 4 groups using both K-means and SOMs.
    Both methods separated the most starburst-like and quiescent galaxies in a similar manner.
    However, they show discrepancies for spectral types between these extreme cases.
    Table~\ref{tab:kmeans_som4} shows the group assignments of the \citetalias{Kinney96} templates in the two methods.
    Templates B, E, S0 and Sa are grouped in the same cluster/neuron in both K-means and $1\times4$-sized SOM.
    K-means grouped the Sb template with the above templates, but the SOM network classified it as a separate group, possibly because it is much more noisy in the ultraviolet range.
    The K-means method clustered the Sc template as a separate group,
    likely due to its distinct features in the 3000--4000~\AA\ range, while the SOM network grouped it with SB5 and SB6.
    Classifying the Sc template in the group with SB5 and SB6 obscures the peak feature in the Sc template.
    However, when the number of SOM nodes is increased (e.g. in Fig.~\ref{fig: 1by22T}), the Sc template separates and occupies a single neuron.
    Another noticeable difference is that K-means clustered the SB1 and SB2 templates in one group and SB3 to SB6 in another, while the SOM network grouped the SB1 to SB4 templates in one neuron and SB5 and SB6 in another.
    SB1 and S2 have similar strong emission lines in the blue, while SB3--6 show H$\alpha$ emission but only weaker emission lines at shorter wavelengths;
    the SB5 and SB6 templates are flatter with a less-pronounced 4000~\AA\ break than the remaining SB templates.
    By comparing the membership of different clusters in Table~\ref{tab:kmeans_som4} we can see that K-means clustering emphasizes the input templates' similarities, while SOM emphasizes the templates' differences.

    \begin{table}
    \centering
    \begin{tabular}{||l|c|c||}
    \hline
    \hline
    \citetalias{Kinney96} template & K-means group & SOM group \\
    \hline
    B                                & C               & 1                         \\
    E                                & C               & 1                         \\
    S0                               & C               & 1                         \\
    Sa                               & C               & 1                         \\
    Sb                               & C               & 2                         \\
    Sc                               & D               & 3                         \\
    SB1                              & B               & 4                         \\
    SB2                              & B               & 4                         \\
    SB3                              & A               & 4                         \\
    SB4                              & A               & 4                         \\
    SB5                              & A               & 3                         \\
    SB6                              & A               & 3                        \\
    \hline
    \end{tabular}
    \caption{Group assignments for \citetalias{Kinney96} template galaxies, using K-means and $1\times4$-sized SOM methods}
    \label{tab:kmeans_som4}
    \end{table}

        \subsubsection{Classifying the galaxy sample}
         \label{sec: 1Dv}

         After training with the templates, we used both K-means and self-organizing maps to classify the 142 galaxies in the \citetalias{Hossein12} sample.
         K-means classification was performed by assigned each galaxy to the K-means cluster with the closest centroid and the results are discussed in Section~\ref{sec: Kmeansvssomvsann}.
        This section examines the results of one-dimensional SOM-based classification in detail.

\begin{figure}
                \begin{subfigure}[b]{0.45\textwidth}
                    \centering
                    \includegraphics[width=\textwidth]{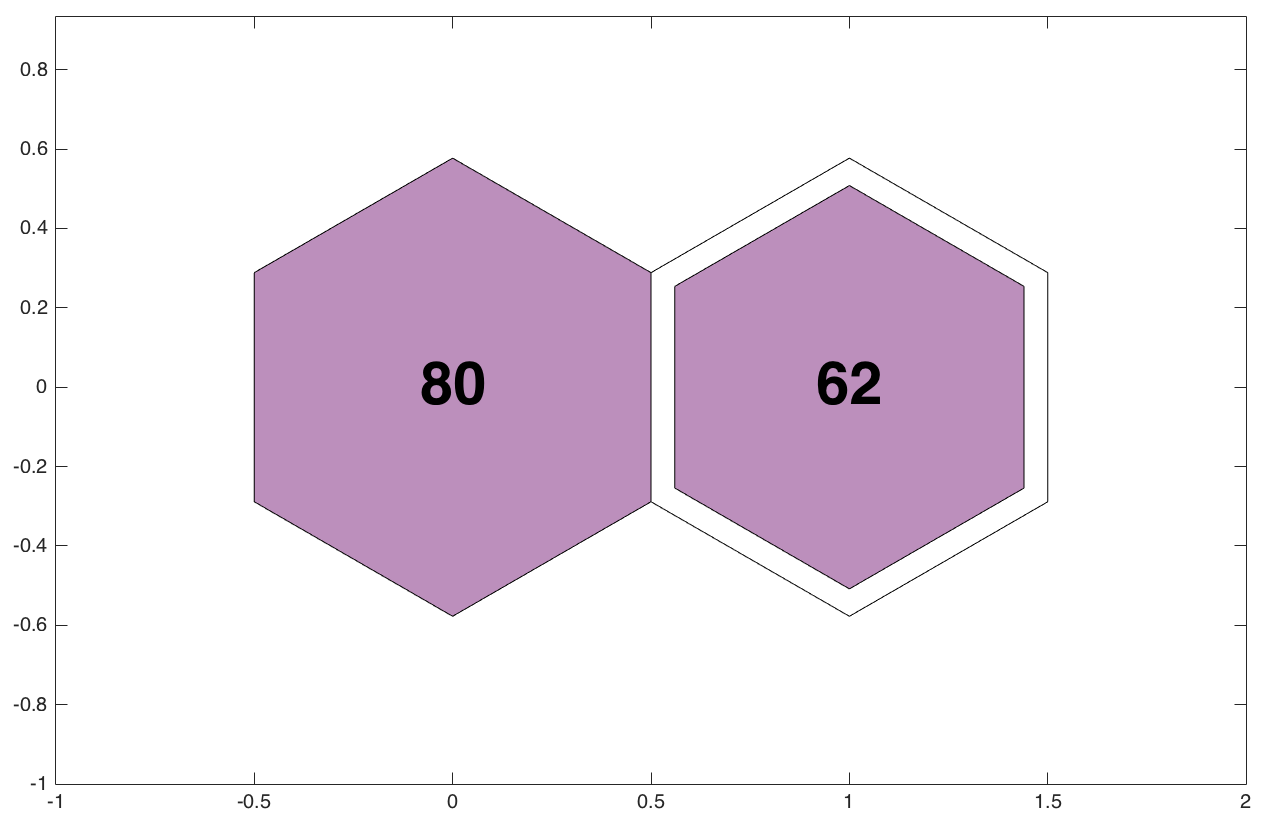}
                \end{subfigure}
                \hfill
                \begin{subfigure}[b]{0.45\textwidth}
                     \includegraphics[width=\textwidth]{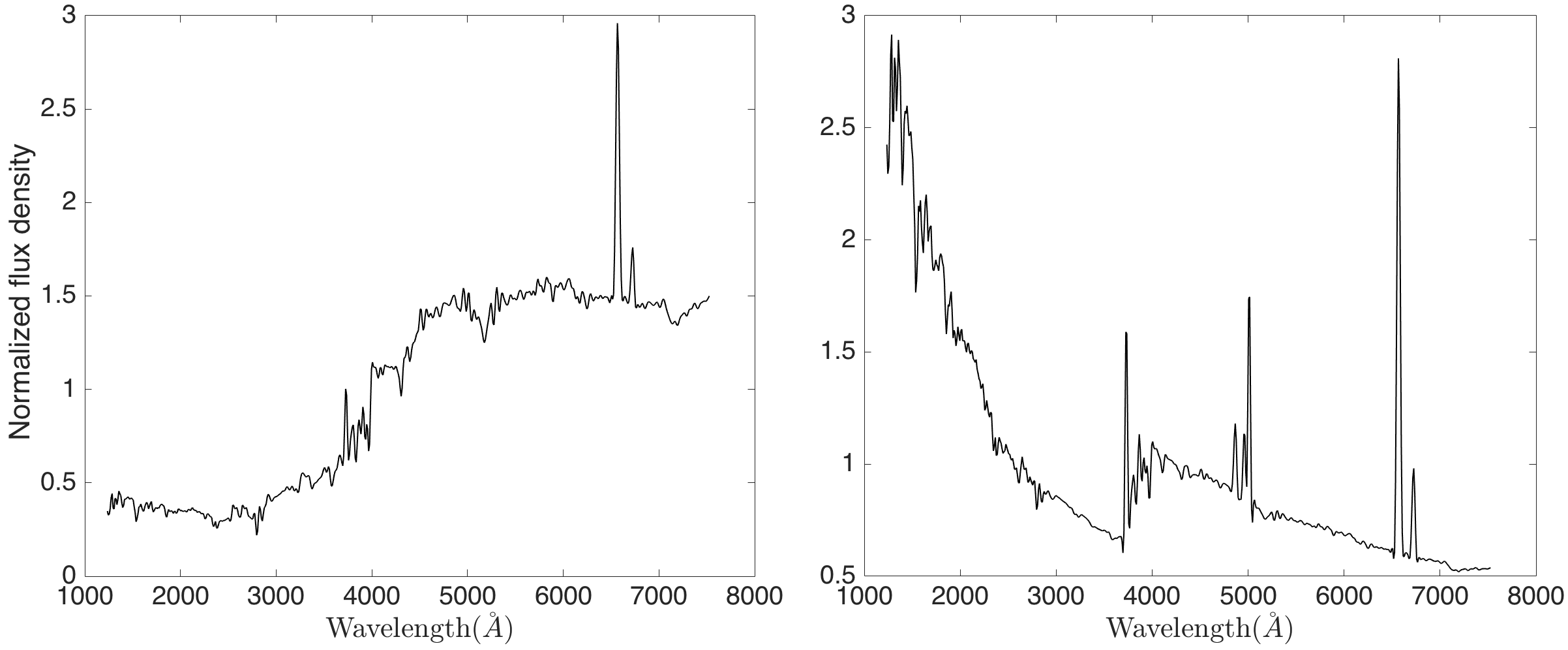}
                \end{subfigure}
                \caption{Classification of fitted galaxy spectra from \citet{Hossein12} using the $1\times2$~network trained from the \citet{Kinney96} templates (Fig.~\ref{fig: 1by2T}). Upper panel: a hit map with the number in each node representing the number of galaxies belonging to that group. In this case 80 means that 80 of the 142 spectra are classified into the quiescent group while the 62 of the 142 spectra are categorized as starburst galaxies. Lower panel: median spectra of the galaxies in each group (flux densities are normalized by dividing spectra by flux density at 4000\AA), quiescent on the left and starburst on the right.}
                \label{fig: 1by2V}
            \end{figure}

            The upper panel in Fig.~\ref{fig: 1by2V} shows the result of this classification using the $1\times2$~network from Fig.~\ref{fig: 1by2T}.
            Eighty of the galaxies have spectral types similar to those of the quiescent galaxies and the spectral types of the rest are similar to the starburst galaxies.
            The lower left panel shows the median spectrum of the 80 galaxies that were classified as quiescent.
            These galaxies are similar to the ones in the left node in the upper panel of Fig.~\ref{fig: 1by2T}:
            the spectra clearly show the 4000\AA~break, one of the signatures of quiescent galaxies.
            The H$\alpha$ emission in the spectrum could be from galaxies with similar spectral types to Sa galaxies, but with stronger emission lines.
            The median spectrum in the lower right panel of Fig.~\ref{fig: 1by2V} shows strong emission lines and bright ultraviolet continuum, indications of a high star formation rate.

            Since in this network galaxies were forced to be divided into a maximum of two groups, the strongest feature in a galaxy's spectrum predominantly decided to which group the galaxy belonged.
            Galaxies with a weak 4000~\AA~break, strong emission lines and ultraviolet upturn are categorized as starbursts while galaxies with a strong 4000~\AA~break are categorized as quiescent.
            Increasing the size of the SOM helps solve the problem of galaxies which have features that are common in both groups.

\begin{figure}
                \begin{subfigure}[b]{0.5\textwidth}
                    \centering
                    \includegraphics[width=\textwidth]{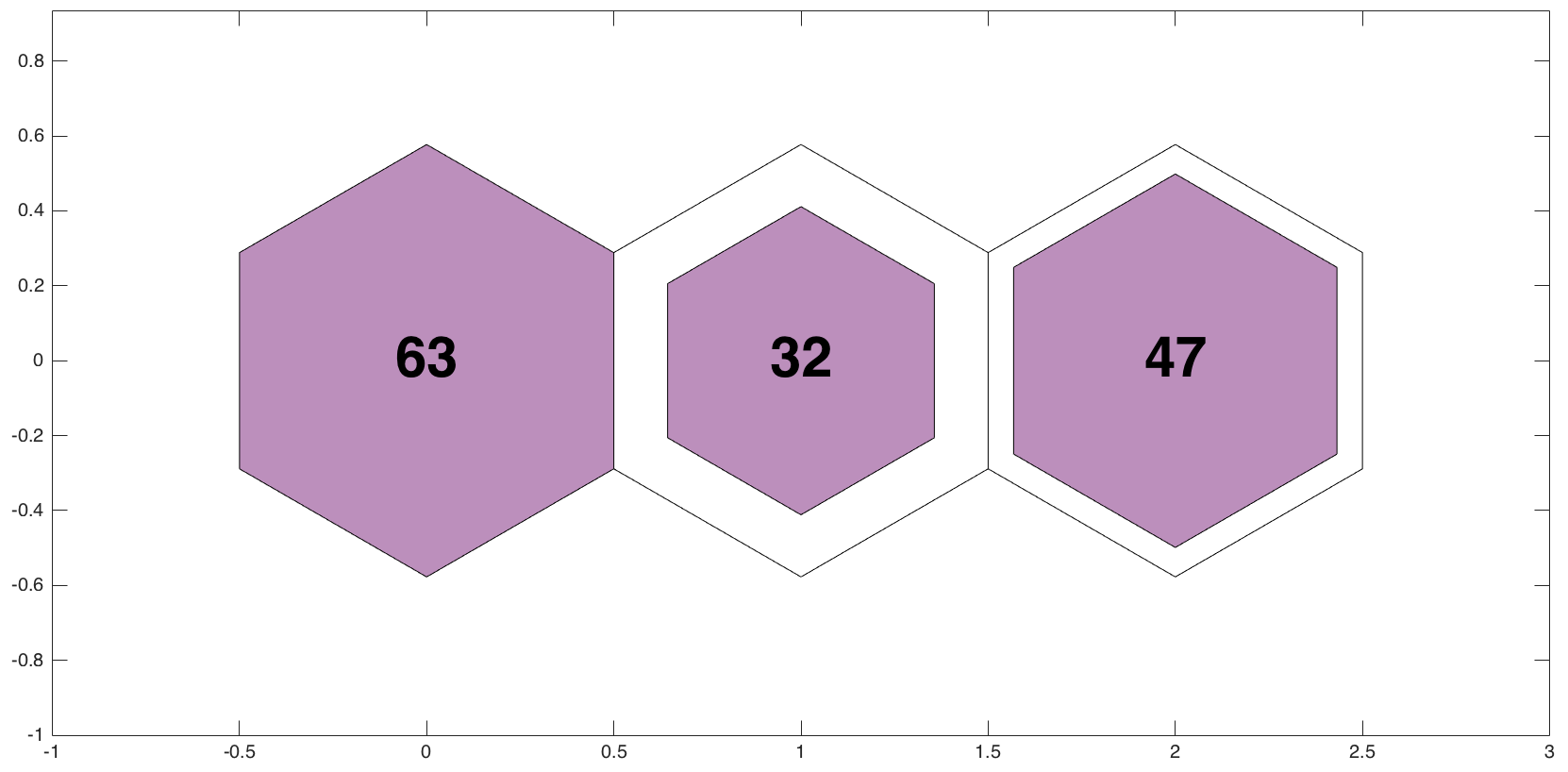}
                \end{subfigure}
                \hfill
                \begin{subfigure}[b]{0.5\textwidth}
                     \includegraphics[width=\textwidth]{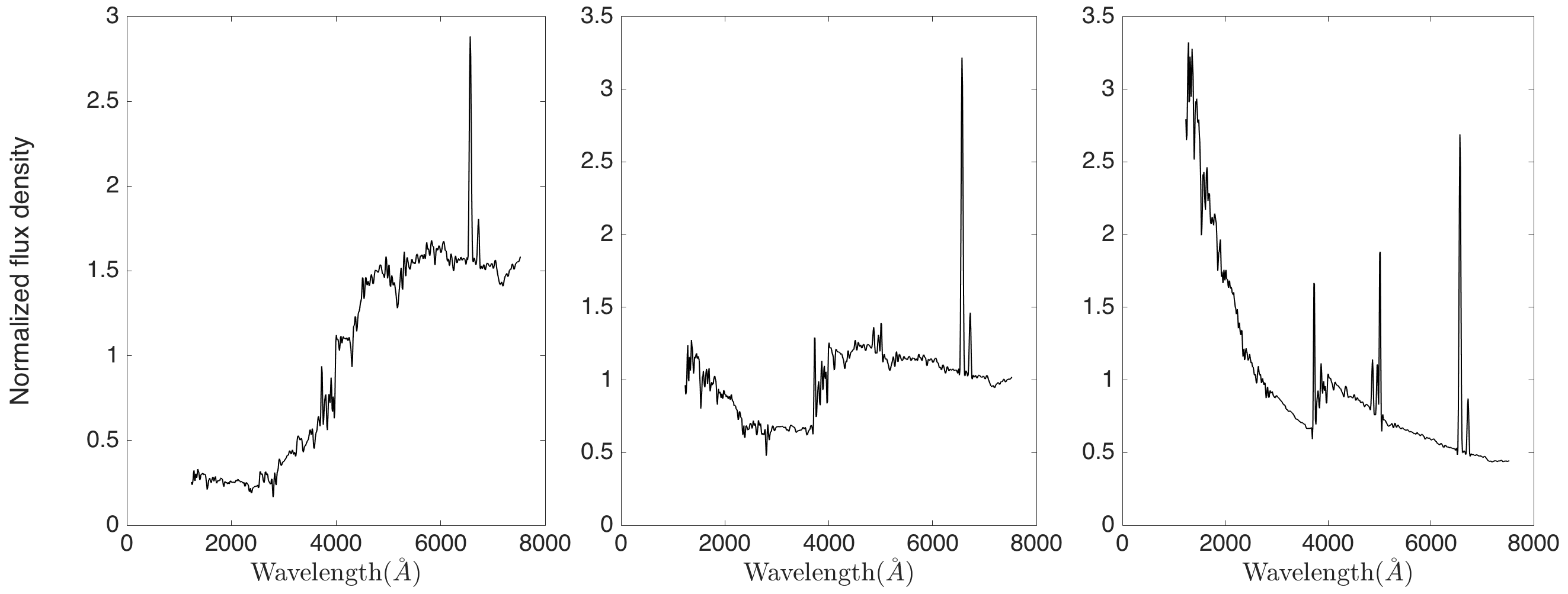}
                \end{subfigure}
                \caption[Classification of fitted galaxy SEDs from \citet{Hossein12} using the $1\times3$~networks]{Same as Fig.~\ref{fig: 1by2V}, but in this figure, we used a network with size of $1\times3$ (Fig.~\ref{fig: 1by3T}) to classify the sample galaxies.}
                \label{fig: 1by3V}
            \end{figure}

            Fig.~\ref{fig: 1by3V} presents the result of classifying the spectra of the galaxies using the $1\times3$~network (from Fig.~\ref{fig: 1by2T}): 66 of the galaxies belong to the quiescent group, and 47 belong to the starburst group.
            However, 29 of the galaxies are similar to starbursts in some, but not all, of their features.
            Galaxies in this group have strong emission lines and are ultraviolet-bright, but they also have a strong 4000~\AA~break, which makes them cluster closer to the quiescent galaxies (middle panel in the lower part of Fig.~\ref{fig: 1by2V}).

\begin{figure*}
                \begin{subfigure}[b]{\textwidth}
                    \includegraphics[width=\textwidth]{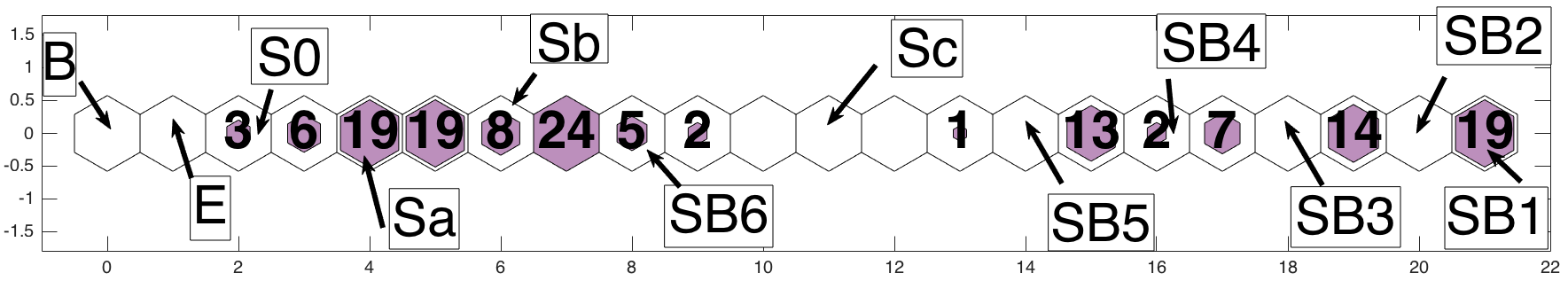}
                    \centering
                \end{subfigure}
                \hfill
                \begin{subfigure}[b]{\textwidth}
                \centering
                     \includegraphics[width=\textwidth]{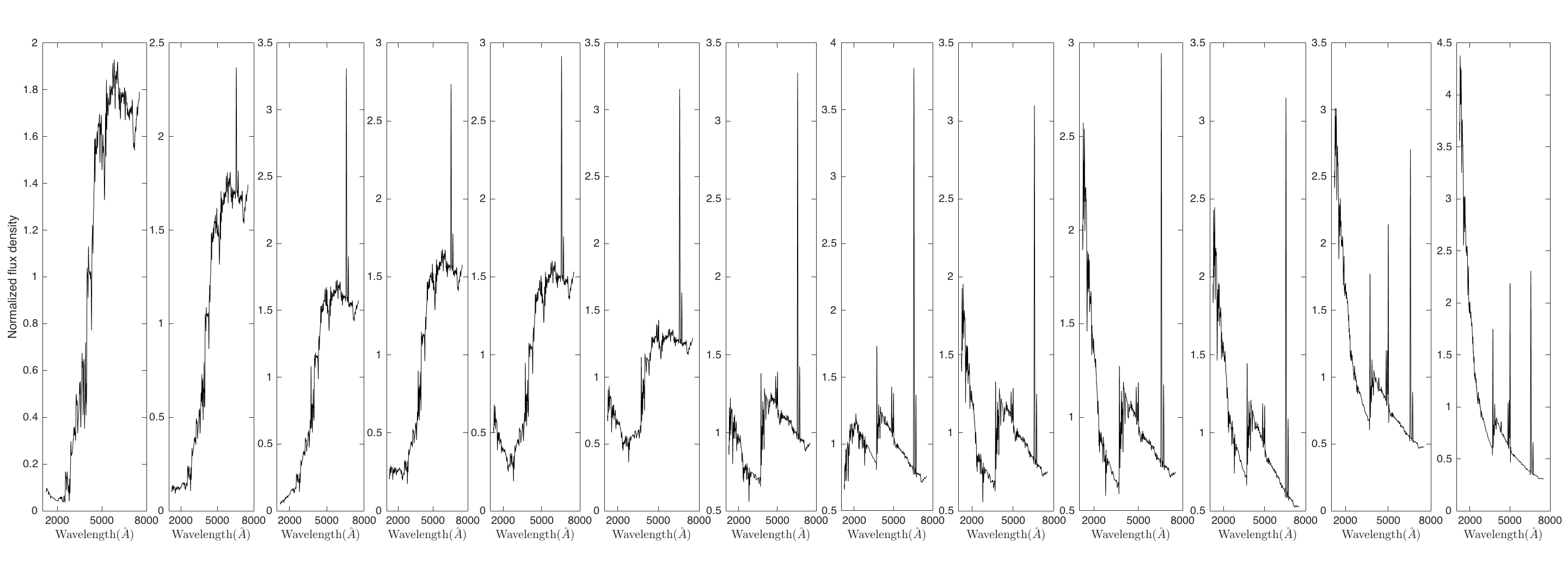}
                \end{subfigure}
                \caption[Classification of fitted galaxy SEDs from \citet{Hossein12} using the $1\times22$~networks]{Same as Fig.~\ref{fig: 1by2V}, but in this figure, we used a network with size of $1\times22$ (Fig.~\ref{fig: 1by22T}) to classify the sample galaxies. In the median spectra of each neuron, we can clearly see the changes from quiescent galaxies to the starburst ones. The 4000\AA\ break weakens from left to right and the emission line strengths increase.}
                \label{fig: 1by22V}
            \end{figure*}

            In Fig.~\ref{fig: 1by22V}, we use the $1\times22$~network to classify the sample galaxies.
            As mentioned in Section~\ref{sec: 1Dt}, in this network size we observed the first separation of the \citetalias{Kinney96} templates into 12 different neurons.
            As in Figs.~\ref{fig: 1by2V} and ~\ref{fig: 1by3V}, the upper panel of Fig.~\ref{fig: 1by22V} shows the number of galaxies (out of the 142) belonging to each neuron in the $1\times22$ SOM.
            The lower panels present the median spectra of the galaxies in each neuron.
            Since galaxies had more space to separate, some of the neurons are left empty.
            Thus, instead of having 22 median spectra in the lower part of Fig.~\ref{fig: 1by22V}, there are only 14.

            Comparing the upper panel of Fig.~\ref{fig: 1by22V} with the lower panel of Fig.~\ref{fig: 1by22T} shows that the occupied neurons are not necessarily the same.
            If a cluster from the \citetalias{Hossein12} sample fills the same neuron as a \citetalias{Kinney96} template, we can conclude that the spectra of galaxies in the cluster are very similar to those of the template.
            Otherwise, we can conclude that the spectra of galaxies in the cluster are (dis)similar to two of the \citetalias{Kinney96} templates.
            For the latter case, the colours establish which template has greater similarity to the \citetalias{Hossein12} cluster.

            The first two neurons in the upper plot in Fig.~\ref{fig: 1by22V} are empty, while these neurons were occupied by galaxies B and E in the trained network.
            We therefore conclude that there are no galaxies in the \citetalias{Hossein12} sample with spectral type similar to that of type B or E.
            There are 3 galaxies in the third neuron, similar to the S0 type.
            19 of the galaxies are in the fifth neuron and similar to the Sa type, while the 6 galaxies in the fourth neuron have spectral type similar to both S0 and Sa galaxies.
            There are 8 Sb type galaxies and 19 galaxies with spectral types similar to both Sa and Sb.
            The following two neurons (eighth and ninth neurons from the left) are on the edge of the quiescent and starburst galaxies.
            In the upper panel in Fig.~\ref{fig: 1by22T}, the colour between these two neurons is black, which shows that these two are very different.
            Therefore, 24 galaxies in the eighth neuron have similar spectra to Sb type galaxies, and the spectra of the 5 galaxies in the ninth neuron are similar to the SB6 galaxies.
            The far right neurons in the network belong to galaxies with spectra comparable to types SB1 and SB2.

    \subsection{Two-dimensional self-organizing maps}
    \label{sec: 2D}
        The one-dimensional networks are great tools to categorize the spectral types of galaxies and monitor the changes of properties of galaxies with category.
        However, each neuron in these networks is limited to connecting to one other neuron in each direction.
        In a 2D network, each neuron can have more than two immediate neighbours, providing a more complete picture of relations between spectral types.
        One of the other key features of SOMs compared to supervised classification is the ability to both interpolate between and extrapolate outside of training set classes.
        This permits a fuller exploration of dataset properties and detection of outliers.

        As described in Section~\ref{sec: method_somz}, one of the main advantages of the SOM is that when the weight of one neuron is adjusted after finding the best matching unit, the weight of the whole map will be changed.
        This quality of the SOMs provides a unique opportunity to analyse 2D networks using two approaches.
        At first, as in the 1D SOMs, we assumed that all the galaxies must have spectra similar to any of the 12 templates in Fig.~\ref{fig: k96}~(i.e. no outliers).
        In this case, we can categorize spectra of other galaxies based on those 12 types.
        In the second approach, we trained networks with the assumption that there might be completely different types of galaxies not encountered before.
        This approach can be used to trace and recognize the outliers, which could include galaxy types that are not present in the \citetalias{Kinney96} templates (e.g. dwarf galaxies, luminous infrared galaxies, or active galactic nuclei).
        To execute the second approach, we used the \textsc{selforgmap} library in \textsc{matlab}.
        Unlike the \textsc{newsom} library in which the ordering step neighbourhood distance is set by default (although in an SOM it should be a free parameter), in \textsc{selforgmap} we can change this value.
        Using the \textsc{selforgmap} code, based on the size of a map and the ordering step neighbourhood distance, we can arrange that the data spread all over the SOM or sit close together, allowing room for outliers.

        To get the best result from both methods, we should have SOMs with ``enough"  neurons to give any new types of galaxies a chance to find their places in the map.
        ``Enough neurons" is a vague statement, but as mentioned in Section~\ref{sec: method_somz} the size of SOMs is arbitrary and depends on the size of the input data and the kind of information needed from the SOMs.
        Since the training dataset (the \citetalias{Kinney96} templates) has 12 members, we assumed a SOM with size of $8\times8$~would be a good start.
        We then increased the size of the SOM to find the highest size suitable for the training sample.
        We noticed that very few changes occurred after the SOMs were grown to a size greater than $12\times12$, so we used this as the maximum size.

        \begin{figure*}
            \begin{subfigure}[b]{0.45\textwidth}
                \centering
                \includegraphics[width=\textwidth]{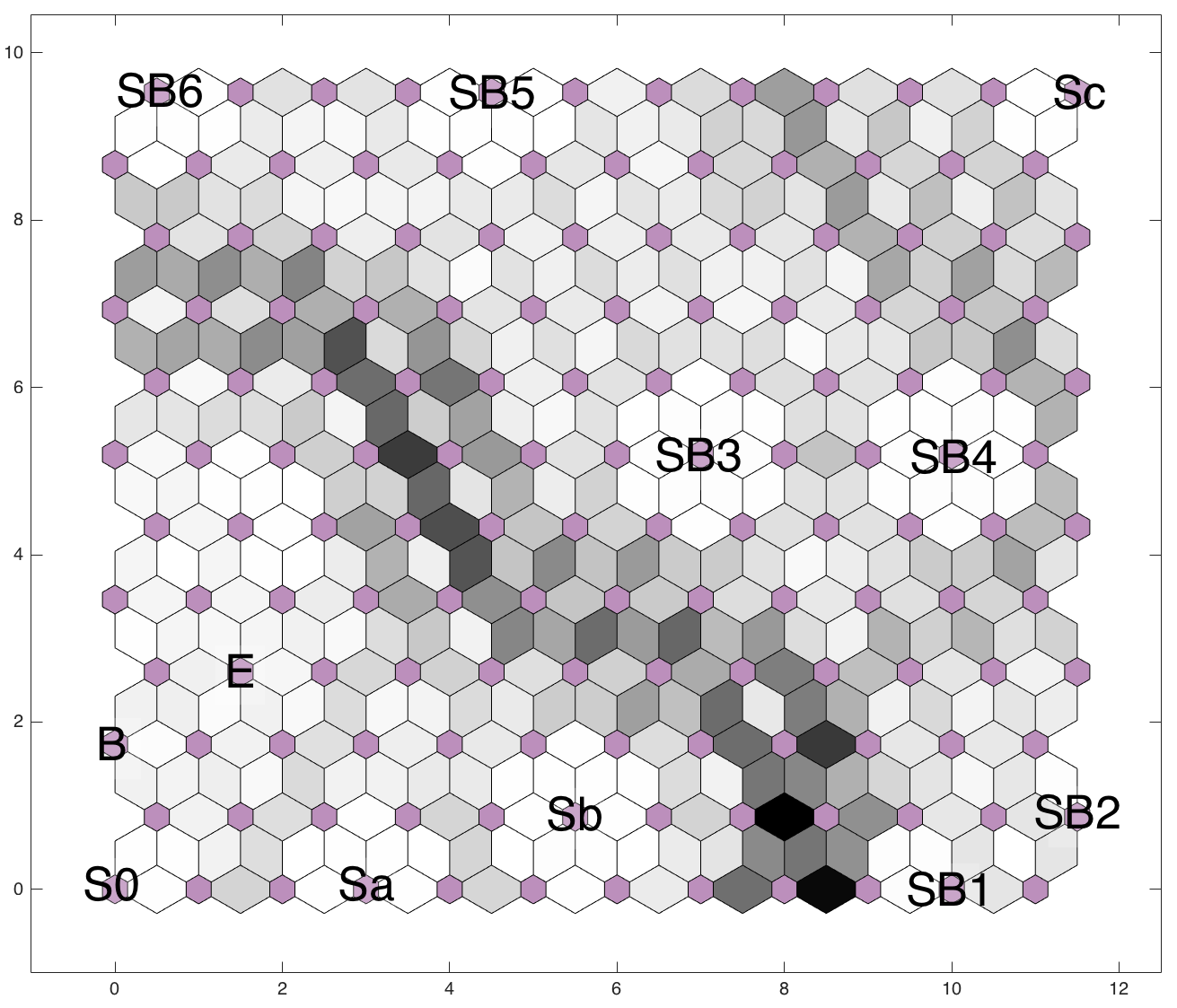}
            \end{subfigure}
            \hfill
            \begin{subfigure}[b]{0.45\textwidth}
                \centering
                \includegraphics[width=\textwidth]{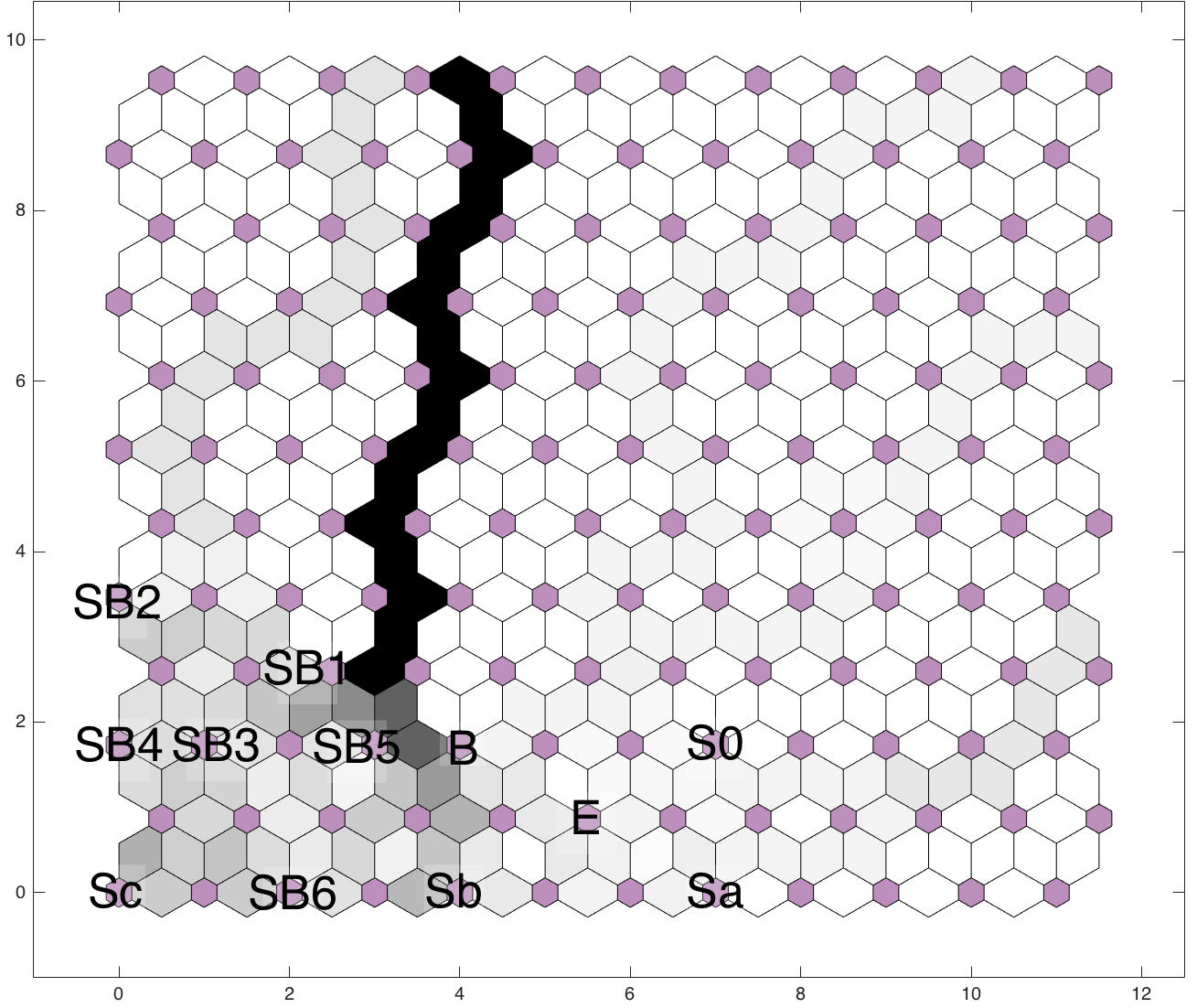}
            \end{subfigure}
            \hfill
            \begin{subfigure}[b]{0.45\textwidth}
                \includegraphics[width=\textwidth]{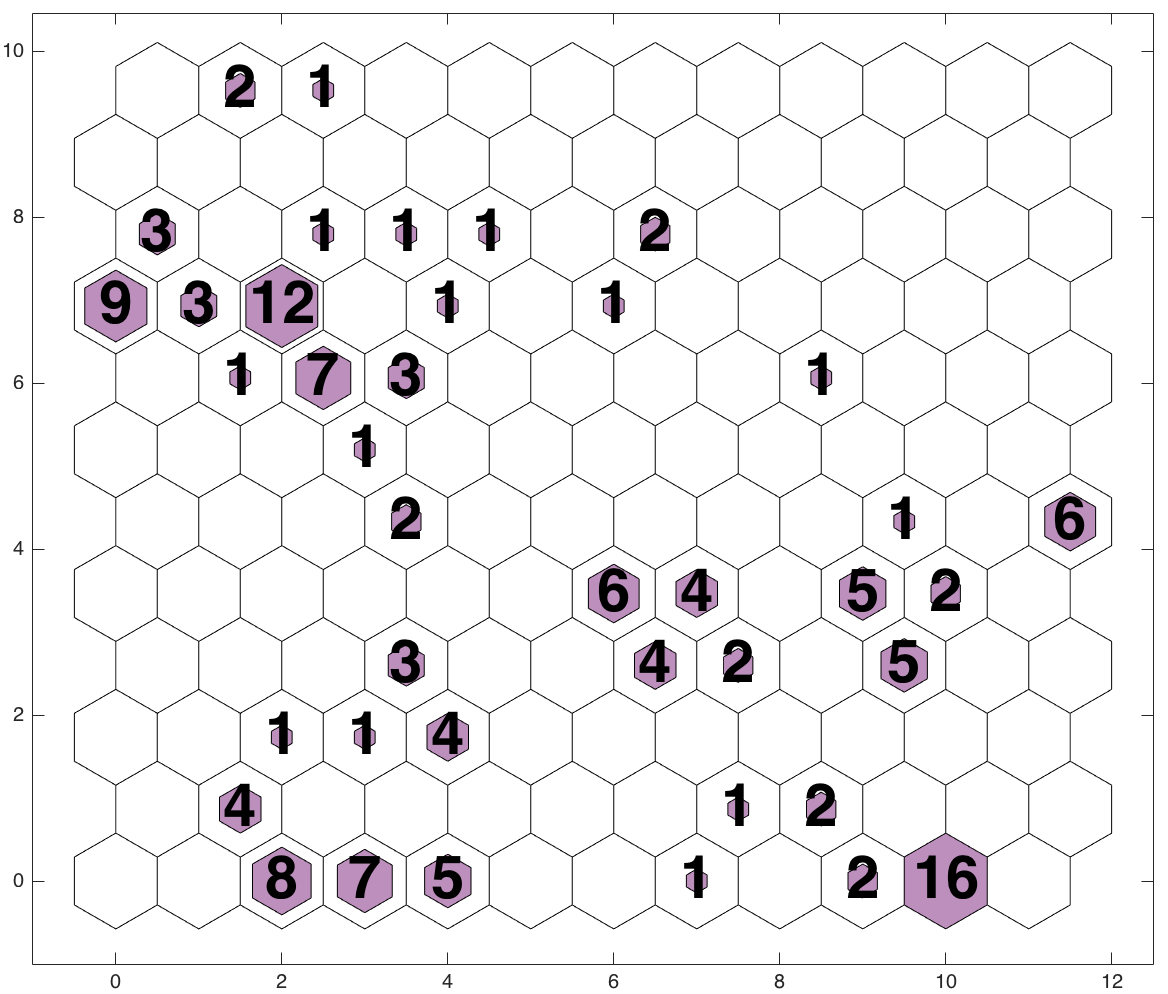}
            \end{subfigure}
            \hfill
            \begin{subfigure}[b]{0.45\textwidth}
                \includegraphics[width=\textwidth]{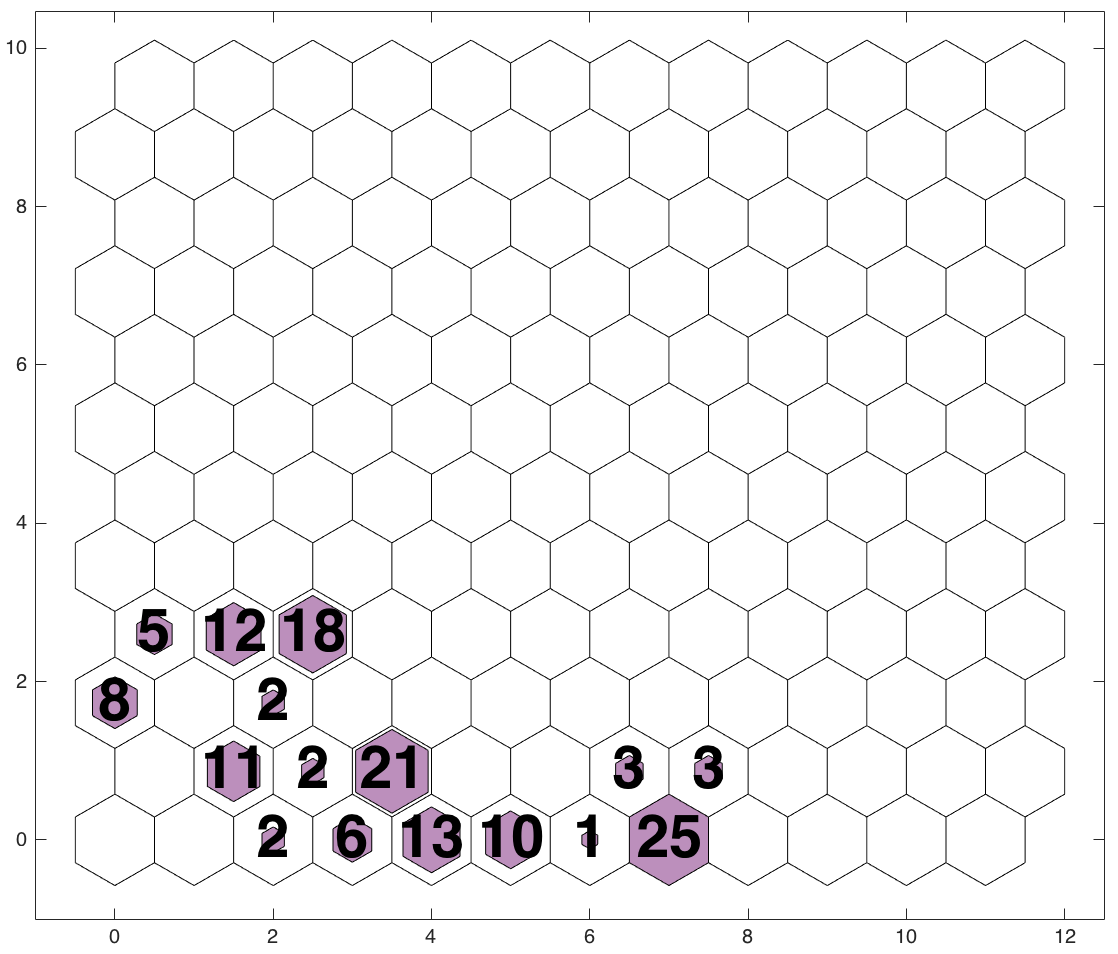}
            \end{subfigure}
            \caption[$12\times12$ two-dimensional self-organizing map results]{$12\times12$~2D SOM results. Upper panel: SOMs trained using the \citet{Kinney96} templates. For training 2D SOMs two different approaches were considered: either only 12 types of galaxies exist (left) or not (right). Lower panel: classifying the galaxies from \citet{Hossein12}, using the trained networks shown in the upper panels. From the lower right panel, we can see that there are no outliers in the galaxies from \citetalias{Hossein12}, and we can use the map in the lower left panel as a final clustering result for the \citetalias{Hossein12} galaxies.}
            \label{fig: 12by12}
        \end{figure*}

        The upper left panel in Fig.~\ref{fig: 12by12} shows the SOM results from the first approach.
        Since we considered that spectra of all galaxies can be categorized using the \citetalias{Kinney96} templates, the galaxies were placed all over the map.
        Using this network to categorize any set of spectra forced the spectra to be either in the same neurons as the \citetalias{Kinney96} templates or in the neurons between them.
        In the map shown in the upper right panel in Fig.~\ref{fig: 12by12}, the \citetalias{Kinney96} templates are in a small region. This provides enough freedom for the spectra of galaxies to be placed everywhere in the map, even far away from the templates.

        In the upper left part of the map in Fig.~\ref{fig: 12by12}, although galaxies have more ways to be separated, they were separated into two main groups.
        There is a distinguishable strip of grey, dark grey and in some cases black colour in the map:
        this strip separates quiescent galaxies from starburst ones.
        The lower map shows that 5 of the neurons on the left side of the strip are full.
        These five neurons are the same as the ones on the left hand side of Fig.~\ref{fig: 1by2T} (quiescent galaxies),
        the only difference being that in this map they have more space to be separated from each other.
        When we use this network to categorize the spectra of galaxies, any galaxy placed on the left hand side of the strip is a quiescent galaxy and any one placed on the right side of the strip is a starburst one.
        The decision about what sub-type of quiescent or starburst galaxy is based upon its position relative to each type in the SOM.
        \citetalias{Hossein12} showed that by masking emission lines (i.e., by ignoring details in spectra) two types of galaxies, Sb and SB6, can be misclassified as one another.
        As mentioned before, few-node, one-dimensional SOMs are a good method to obtain a big picture of the problem under study with less emphasis on fine detail.
        For example, the SB6 category is placed near Sb galaxies in the 1D network in Fig.~\ref{fig: 1by22T}.
        However, in the upper left panel of Fig.~\ref{fig: 12by12}, we can see that the Sb and SB galaxies are completely separated.
        This shows that more details of spectral shapes are considered by 2D networks, which is one of the main advantages of  using 2D maps over 1D ones.
       In this case, the two dimensions of the self-organizing map correspond approximately to quiescent versus star-forming (left to right) and less to more extinction (bottom to top).

        As in the other map, in the upper right panel of Fig.~\ref{fig: 12by12}, galaxies are generally divided into two main groups.
        The border between the quiescent galaxies and the starburst ones is the black strips in the middle of the map, ending with the bright grey colour at the bottom of the map in the fifth neuron.
        In this network, neurons in the right side of the strip represent the quiescent galaxies and the neurons in the left side represent the starburst ones.
        When categorizing a new set of spectral types, if the new spectra are similar to the \citetalias{Kinney96} templates all of them will be placed in the bottom of the map, but if the new spectra represent different types of galaxies, they would sit in any other neurons in the map.
        In large datasets, one can easily use this network to figure out whether there is any of new type of spectra (or any outliers) in the datasets.
        Since the networks are already available, this procedure should be quick (from a few seconds to a few minutes) and easy for big datasets.

        We used both 2D networks to categorize the \citetalias{Hossein12} galaxies and the results are shown in the lower panels of Fig.~\ref{fig: 12by12}.
        In the lower right map in Fig.~\ref{fig: 12by12}, all galaxies are placed in the bottom part of the map and we can conclude that in the \citetalias{Hossein12} sample there are no outliers or spectra with very different from the \citetalias{Kinney96} templates.
        The lower left map of Fig.~\ref{fig: 12by12} shows the \citetalias{Hossein12} galaxies categorized based on the network in the upper left of Fig.~\ref{fig: 12by12}.
        Comparing this categorization with the 1D one from Fig.~\ref{fig: 1by22V}, we can see that only 23 galaxies correspond exactly to \citetalias{Kinney96} types.
        Using 2D maps results in categorizing galaxies into more intermediate types than in 1D maps.
        In both lower maps in Fig.~\ref{fig: 12by12}, most galaxies are in the quiescent side of the SOM, which was predictable from the results in Section~\ref{sec: 1Dv}.
        This does not necessarily imply that in general there are more quiescent galaxies at higher redshift but could be a selection effect for galaxies that had more reliable redshift estimates.

    Although for ease of presentation this paper first discusses 1D networks and continues to 2D networks, we suggest that users of SOMs for galaxy spectral classification should use 2D networks first.
         These can be used as a first step with networks similar to the upper left map in Fig.~\ref{fig: 12by12} to identify outliers in the sample.
        After finding and removing outliers in the input data, the use of 1D and 2D maps is complementary.
        One-dimensional maps are helpful for exploring the general and ordered behaviour of the data, while 2D maps can show a more detailed picture.

\subsection{Comparing classification methods}
    \label{sec: Kmeansvssomvsann}

    The one and two-dimensional self-organizing map classifications are compared briefly above.
    The 2-D maps allow galaxies to be members of more intermediate classes, and this is a general advantage of SOMs compared to the artificial neural network method used by \citetalias{Hossein12}: in the SOM method galaxies do not need to exactly match template spectra.
    \citetalias{Hossein12} found that the ANN method could only match 105 out of the 142 galaxies directly to the \citetalias{Kinney96} templates; classifying the remaining 37 galaxies required combining template spectra.
    Below we compare various other classification methods with the results of the 1-D self-organizing maps using the metrics described in Section~\ref{sec:metrics}.
    Because the 2-D SOMs subdivide the sample into a larger number of groups, each containing a small number of objects, they are not directly comparable to the other methods and are not included here.

    \begin{table}
    \centering
    \begin{tabular}{ll}
    \hline
    \hline
    Method & chi-squared agreement score \\
    \hline
    K-Means                          & 124  \\
    $1\times4$-sized SOM             & 103   \\
    $1\times12$-sized SOM            & 111.5   \\
    $1\times22$-sized SOM            & 79   \\
    supervised ANN                   & 52    \\
    \hline
    \end{tabular}
    \caption{Chi-squared agreement score for different methods}
    \label{tab:class_compare}
    \end{table}

       To compare results from SOM, K-means, and supervised ANN, we first consider how well these methods matched a simple chi-squared fitting classification.
        Matching galaxy spectra to templates using chi-squared, we found that not all of the templates had matches among the galaxy sample: no galaxy spectra were best-matched to the B, E, or Sc templates.
        This is consistent with the results from the self-organized maps.
        Summing the chi-squared agreement score over all galaxies in the sample allows us to compare how well each technique matched the results of chi-squared fitting.

        The results, given in \autoref{tab:class_compare}, show that the K-means method and $1\times12$-sized SOM have the most similarity with minimum chi-square fitting.

        Summing the chi-squared agreement score over all methods for a single galaxy shows how well the methods agree for a particular spectrum.
        For example, if the total score of a galaxy is 5, it means that all techniques classified the galaxy into the same group.
        Comparing the scores, we found that 12 per cent of galaxies scored 5, 36 per cent of galaxies had a score greater than 4, and 56 per cent had a total score of 3 or more.
        The score-5 galaxies classified similarly by all methods were primarily (60 per cent) in the Sa template group.
        The score $\geq 3$ galaxies were predominantly SB1, Sa, and Sb type (32, 28, 18 per cent, respectively).
        The different methods show the most agreement for galaxies similar to the SB1 and Sa templates, with less agreement on the high-extinction starburst types (SB5, SB6).

   \begin{table}
    \centering
    \begin{tabular}{ll}
    \hline
    \hline
    Method & Silhouette score \\
    \hline
    K-Means                          & 0.549 \\
    $1\times4$-sized SOM             & 0.371   \\
    $1\times12$-sized SOM            & 0.302   \\
    $1\times22$-sized SOM            & 0.249   \\
    chi-squared fitting              & 0.221 \\
    supervised ANN                   & 0.142    \\
    \hline
    \end{tabular}
    \caption{Silhouette scores for different methods}
    \label{tab:sil_score}
    \end{table}

A final comparison between the classification methods is made with the silhouette score  \citep{rousseeuw87}, given in \autoref{tab:sil_score}.
The highest score among our classification was for K-means; however
as discussed in Section~\ref{sec:metrics}, the silhouette score usually declines as the number of groups increases, so we cannot conclude from score alone that K-means is the superior method.
However, the $1\times12$-sized SOM, chi-squared fitting, and supervised neural network all have either 8 or 9 classes.
In this case we can compare scores directly and conclude that the self-organized map yields groups with more internal similarity.
This is in contrast to the comparison between K-means and self-organized maps for the templates alone (see Section~\ref{sec: Kmeansvssom}) where K-means clustering  produced groups more similar to each other, and emphasizes that the exact data used in a clustering process can strongly affect the outcome.

 \subsection{Properties of spectral groups}

       \begin{figure*}
            \centering
            \includegraphics[width=\textwidth]{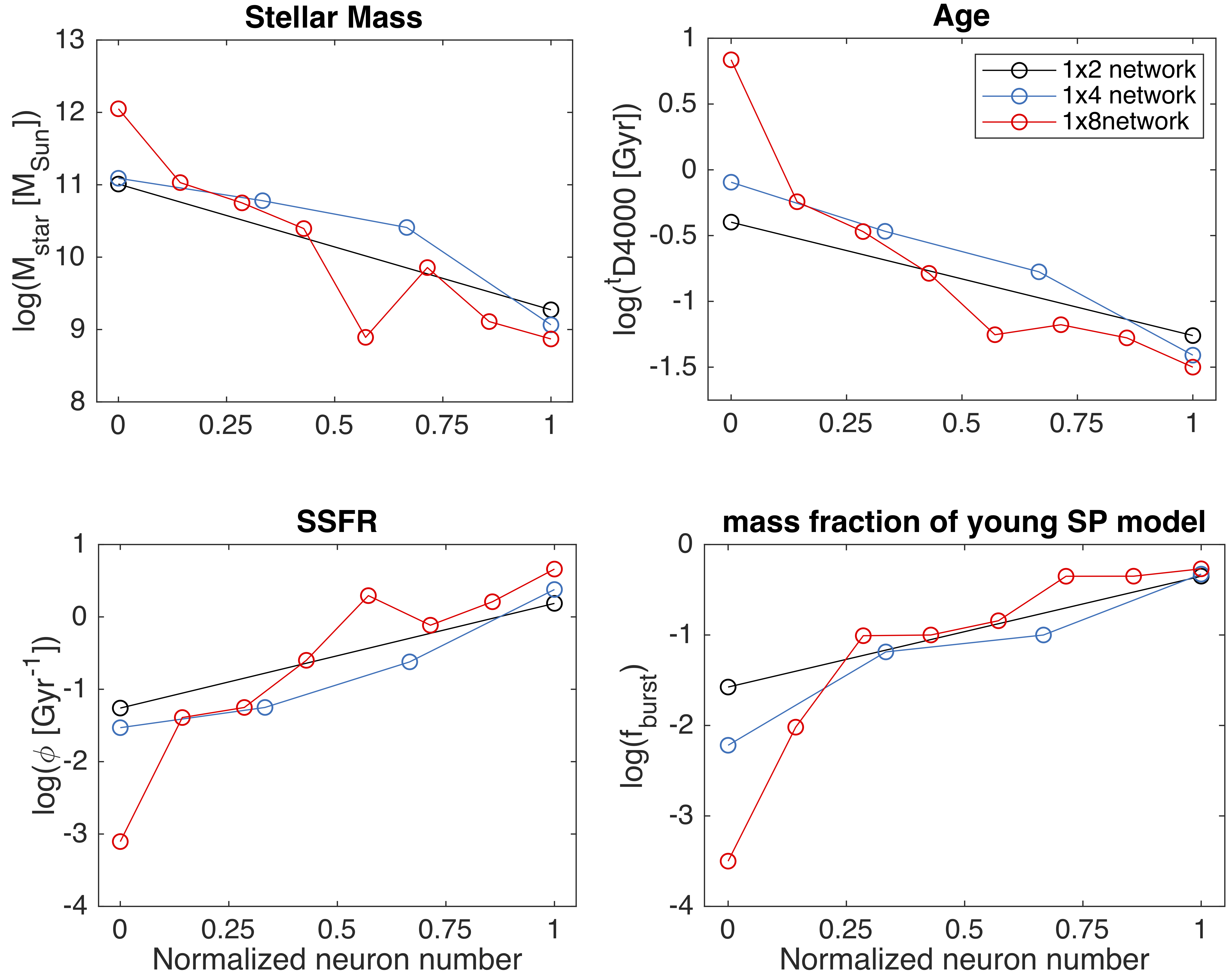}
            \caption[The median of four properties of galaxies in three networks]{Comparing the median of four properties of galaxies in each node in $1\times2$ (black circles), $1\times4$ (blue diamonds), and $1\times8$ (red triangles) networks.}
            \label{fig: props}
        \end{figure*}

       To check whether the classifications described above are meaningful, in this section we examine the relations between median properties of the galaxies in each group.
          Fig.~\ref{fig: props} shows the median of stellar mass, age, specific star formation rate (sSFR; star formation rate per stellar mass), and $f_\mathrm{burst}$ of the galaxies in each neuron in the $1\times2$, $1\times4$, and $1\times8$ self-organized map networks.
        In all plots, the horizontal axis is the number of the neurons divided by the size of the network.
        As shown in Fig.~\ref{fig: props}, in all three networks, stellar mass and age decrease while sSFR and $f_\mathrm{burst}$ increase as the type changes from quiescent galaxies to starburst.
       Separating galaxies based on spectral types also leads to a separation in properties derived (via {\em CIGALE}) from the SEDs, as expected since the spectral types are also based on {\em CIGALE} fitting.

\begin{figure*}
\includegraphics[width=\textwidth]{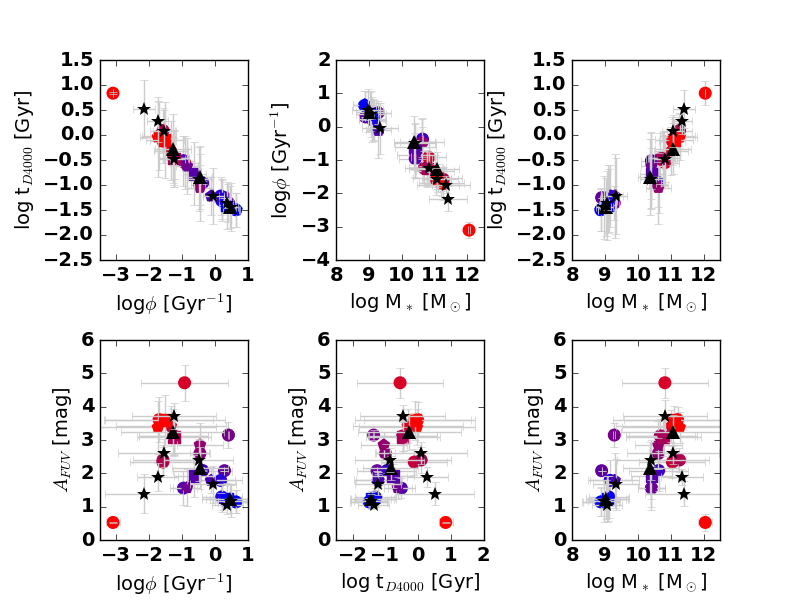}
        \caption{
        From top to bottom: relations between age (t$_{\rm {D4000}}$) and sSFR ($\phi$), sSFR and stellar mass, age (t$_{\rm {D4000}}$) and stellar mass, and A$_{\rm {FUV}}$ with sSFR, age (t$_{\rm {D4000}}$) and stellar mass.
        Circles, pentagons, and squares show the median values of these properties of galaxies in the $1\times22$, $1\times 12$ and $1\times4$ SOM networks;
        triangles show the median value of properties for each K-means cluster (using $K=4$) and stars show median values for groups from \citetalias{Hossein12}.
        Error bars show standard deviations of the group properties.
        Colours show the galaxy types in the SOM classification.
        The lightest grey (most pure red in the online version) colour shows E type galaxies and the black (most blue in the online version) indicates the SB1 galaxies.
        The shading of other colours between light grey and black (red and blue in the online version) measures the location of galaxies along the SOM from  quiescent to starburst.}
        \label{fig: props_vs_props}
        \end{figure*}

        {\em CIGALE} has various models to derive the properties of galaxies.
        Through these models, some of the properties are already known to be correlated with each other, e.g., stellar mass and star formation rate.
        \citetalias{Noll09} studied other relations between properties with no direct correlation in the models in a sample of SINGS galaxies.
        They found a tight correlation between sSFR and t$_{\rm {D4000}}$, which suggests that younger stellar population correlates with higher SFR.
        They also found correlations between stellar mass and SFR, and stellar mass and t$_{\rm {D4000}}$.
        Since in the {\em CIGALE} code, stellar mass is a free parameter, \citetalias{Noll09} argued that any stellar-mass-related correlation must be astrophysically meaningful.
        They also studied relations between the attenuation at 1500 \AA~(A$_{\rm {FUV}}$) and sSFR, age, and stellar mass and did not find any correlation.
        \citetalias{Hossein12} replicated the upper plots in Fig.~\ref{fig: props_vs_props}, and found a tighter correlation than in the \citetalias{Noll09} results.

        To compare our classifications with previous work, Fig.~\ref{fig: props_vs_props} shows plots similar to those shown by both \citetalias{Noll09} and \citetalias{Hossein12}.
        The points represent the median values of the properties of the galaxies in each group, with different symbol types representing classification method and error bars indicating the standard deviation of those properties.
       The colour scale represents neuron number in the SOM method, shading from quiescent to starburst.
        The medium grey (purple in the online version) point in the middle of the black (blue in the online version) points in Fig.~\ref{fig: props_vs_props} corresponds to galaxies with spectra similar to type Sc: the shape of their spectra indicates that they are old galaxies with high sSFR.

        Overall, Fig.~\ref{fig: props} and the upper panels in Fig.~\ref{fig: props_vs_props} show the same trends as noted by \citetalias{Noll09} and \citetalias{Hossein12}:
        galaxies with lower mass tend to be younger and more active.
        The median group properties as derived by all of the methods show high Pearson correlation coefficients ($r^2\gtrsim 0.9$) between the quantities shown.
        The ordering of groups in the SOM maps also corresponds to an ordering in age:
        from older to younger galaxies, the colour of the points changes from light grey to black (red to blue in the online version).

        \citetalias{Noll09} studied relations between A$_{\rm {FUV}}$ and other properties of galaxies and found that the attenuation had no dependence on the specific star formation or age.
        In contrast to \citetalias{Noll09}, we find a correlation between A$_{\rm {FUV}}$ and these two parameters, shown in the lower panels of Fig.~\ref{fig: props_vs_props}.
        Although weaker then the correlations above, Pearson correlation coefficients are higher for the SOM groups than for other methods, particularly for the $1x12$-sized SOM.
        The slightly different distribution of this particular classification (see Fig.~\ref{fig: 1by12T}) classifies the most quiescent objects as similar to slightly more active galaxies, lessening the turnover in properties as A$_{\rm {FUV}}$ increases.

        The general sense of the correlation between A$_{\rm {FUV}}$ and sSFR in Fig.~\ref{fig: props_vs_props} is similar to the trend found by \cite{Dale07}.
        They used IR luminosity to UV luminosity ratio (L$_{IR}$/L$_{FUV}$) as a measure of A$_{\rm {FUV}}$ and compared it with sSFR for all 75 galaxies in the SINGS survey.
        They found that for quiescent galaxies (E, S0 and S0/a), L$_{IR}$/L$_{FUV}$ (or A$_{\rm {FUV}}$) correlates with sSFR, and for spiral galaxies, there is an anticorrelation between L$_{IR}$/L$_{FUV}$ and sSFR.
        Since our sample has no E-type galaxies, we cannot confidently claim the same relation between A$_{\rm {FUV}}$ and sSFR.
        However, Fig.~\ref{fig: props_vs_props} does indicate an increase in A$_{\rm {FUV}}$ with increasing sSFR for the quiescient types and a decrease with increasing sSFR for the starburst types.
        Both \cite{Dale07} and \citetalias{Noll09} argued that these apparent trends can be a result of the dependence of star formation history on L$_{IR}$/L$_{FUV}$.
        Whether this dependence is real or not, our results here show that using self-organizing maps can separate spectra of galaxies in such a way that the characteristics of each of these groups are in agreement with the general picture of galaxy evolution.

       One of the key features of SOMs is the ability to not only classify objects but to provide an {\em ordered} classification.
       K-means clustering and supervised ANNs can sort galaxy spectra into classes but do not naturally order them. Ordering is important in the case where the physical causes of the differences between classes/clusters are unknown. Average properties of objects within clusters (possibly from data not used for classification) can be used to test hypotheses about the physical differences.
       Figures~\ref{fig: props} and~\ref{fig: props_vs_props} show that the SOM ordering also corresponds to the ordering of the physical properties of the galaxy models used to make the classified spectra.
       The median properties of the spectral groups change on going from one end of the SOM to the other.
       Therefore, by using the SOM method, we can study not only how galaxies group into spectral clusters, but the trends between the clusters as well.

       \section{Summary}
       \label{sec: summary_SOMZ}

           Self-organizing maps can be used to classify celestial objects (e.g. stars, quasars, spectra of galaxies, light curves, etc.).
           In this work we presented a detailed comparison between different types of self-organizing maps and other methods used to classify galaxy spectra.
           Based on our experience we suggest some general guidelines for use of SOMs in galaxy spectral classification.
           If a broad and general classification is required, networks can have one dimension with a few neurons.
           If one needs more detailed classifications, a higher number of neurons should be used.
           Since self-organizing maps do not include the uncertainty of input parameters, sometimes too much attention to detail can cause problems in classifications.
           When using the SOM method, one should consider whether small differences between objects are physically meaningful when separating two groups from each other.

           We used SOMs to classify the template spectra of \citetalias{Kinney96}, made from galaxies with known morphological type, and created networks with different uses.
           By varying the size of the networks, we found the relative similarity between the \citetalias{Kinney96} template classes, which can be roughly ordered in one dimension by their amount of star formation.
            A one-dimensional network with 22 neurons was needed in order to
           separate all 12 \citetalias{Kinney96} spectra; we concluded that \citetalias{Kinney96} types B and E, and types SB1 and SB2, are very similar to each other.
           Two-dimensional networks allow more freedom in the galaxy spectral classification.
           Training $12\times 12$ neurons with the \citetalias{Kinney96} spectra, we found that the two dimensions of the resulting self-organized maps corresponded roughly to strength of star formation and amount of extinction. The one-dimensional self-organizing map ordering combined these two properties, with the highest-extinction template SB6 appearing between two less-extincted templates (Sb and Sc).

           A sample of 142 high-redshift galaxy spectra from \citetalias{Hossein12} was classified by the trained networks.
         This particular sample of high-redshift galaxies is well-described by the range of the \citetalias{Kinney96} templates although many of the spectra fall in between template classes.
            In the two-dimensional network, only 23 galaxies occupied exactly the same neurons as the \citetalias{Kinney96} template spectra (in one-dimensional networks this number was 56).
           The freedom of having in-between types is one of the main differences between supervised and unsupervised artificial neural networks: the supervised training method used by \citetalias{Hossein12} could not classify 37 out of 142 spectra in galaxy sample as matching one of the same set of templates.

           Comparing different classification methods, we found that spectra classified similarly by all methods were well-matched to quiescent templates (\citetalias{Kinney96} Sa and Sb).
           Spectra with emission lines or ultraviolet emission showed more variable classification results.
            As measured by a chi-squared agreement score,
            classification produced by self-organized maps showed closer correspondence to the classification from a chi-squared match than results from the supervised ANN method.
           One-dimensional self-organized-map based classification resulted in a
           higher silhouette score  than classification using chi-squared minimization or the supervised neural network, indicating that the galaxies classed together by the SOM had more similar spectra.

       The relations between group-averaged properties of the sample galaxies using the SOM-based classification were found to be consistent with those from previous studies.
           The self-organizing map method can order the spectra in such a way as to also match the order along the age/star formation rate and specific star formation rate-stellar mass relations.
           Such an ordered classification does not naturally arise from other methods such as K-means clustering and supervised artificial neural networks.
           Ordered classification can be used to investigate underlying physical causes of changes in properties between groups and for this reason we conclude that self-organized maps can be a highly useful tool in exploratory analysis of astronomical spectra.

\section*{ACKNOWLEDGMENTS}
The authors thank the referee for thorough and insightful comments which helped us to improve the work.
The authors thank S. Lianou, A. Tammour, S.C. Gallagher, R.G. Abraham, M. Daley, M. Gorski and A. Sigut for their useful comments.
S.R. and P.B. also acknowledge research support from the Natural Sciences and Engineering Research Council of Canada.

\bibliographystyle{apalike}
\bibliography{ref_mining_h.bib}

\begin{thebibliography}{}

\bibitem[\protect\citeauthoryear{{Abazajian}, {Adelman-McCarthy},
  {Ag{\"u}eros}, {Allam}, {Allende Prieto}, {An}, {Anderson}, {Anderson},
  {Annis}, {Bahcall} \& et al.}{{Abazajian} et~al.}{2009}]{Abazajian09}
{Abazajian} K.~N.,  {Adelman-McCarthy} J.~K.,  {Ag{\"u}eros} M.~A.,  {Allam}
  S.~S.,  {Allende Prieto} C.,  {An} D.,  {Anderson} K.~S.~J.,  {Anderson}
  S.~F.,  {Annis} J.,  {Bahcall} N.~A.,    et al. 2009, \apjs, 182, 543

\bibitem[\protect\citeauthoryear{{Allen}, {Hewett}, {Richardson}, {Ferland} \&
  {Baldwin}}{{Allen} et~al.}{2013}]{Allen13}
{Allen} J.~T.,  {Hewett} P.~C.,  {Richardson} C.~T.,  {Ferland} G.~J.,
  {Baldwin} J.~A.,  2013, \mnras, 430, 3510

\bibitem[\protect\citeauthoryear{{Andreon}, {Gargiulo}, {Longo}, {Tagliaferri}
  \& {Capuano}}{{Andreon} et~al.}{2000}]{Andreon00}
{Andreon} S.,  {Gargiulo} G.,  {Longo} G.,  {Tagliaferri} R.,    {Capuano} N.,
  2000, \mnras, 319, 700

\bibitem[\protect\citeauthoryear{{Balastegui}, {Ruiz-Lapuente} \&
  {Canal}}{{Balastegui} et~al.}{2001}]{Balastegui01}
{Balastegui} A.,  {Ruiz-Lapuente} P.,    {Canal} R.,  2001, \mnras, 328, 283

\bibitem[\protect\citeauthoryear{{Ball}, {Loveday}, {Fukugita}, {Nakamura},
  {Okamura}, {Brinkmann} \& {Brunner}}{{Ball} et~al.}{2004}]{Ball04}
{Ball} N.~M.,  {Loveday} J.,  {Fukugita} M.,  {Nakamura} O.,  {Okamura} S.,
  {Brinkmann} J.,    {Brunner} R.~J.,  2004, \mnras, 348, 1038

\bibitem[\protect\citeauthoryear{Beale, Hagan \& Demuth}{Beale
  et~al.}{2013}]{matlabtolbox}
Beale M.,  Hagan M.,    Demuth H.,  2013, MATLAB Neural Network Toolbox User's
  Guide, MathWorks, Version 8.0.1 (Release 2013a)

\bibitem[\protect\citeauthoryear{{Bershady}, {Jangren} \&
  {Conselice}}{{Bershady} et~al.}{2000}]{Bershady00}
{Bershady} M.~A.,  {Jangren} A.,    {Conselice} C.~J.,  2000, \aj, 119, 2645

\bibitem[\protect\citeauthoryear{Bishop}{Bishop}{2006}]{bishop06}
Bishop C.~M.,  2006, Pattern Recognition and Machine Learning (Information
  Science and Statistics).
Springer-Verlag New York, Inc., Secaucus, NJ, USA

\bibitem[\protect\citeauthoryear{{Boersma}, {Bregman} \&
  {Allamandola}}{{Boersma} et~al.}{2014}]{Boersma14}
{Boersma} C.,  {Bregman} J.,    {Allamandola} L.~J.,  2014, \apj, 795, 110

\bibitem[\protect\citeauthoryear{{Brett}, {West} \& {Wheatley}}{{Brett}
  et~al.}{2004}]{Brett04}
{Brett} D.~R.,  {West} R.~G.,    {Wheatley} P.~J.,  2004, \mnras, 353, 369

\bibitem[\protect\citeauthoryear{{Carrasco Kind} \& {Brunner}}{{Carrasco Kind}
  \& {Brunner}}{2014}]{Kind14a}
{Carrasco Kind} M.,  {Brunner} R.~J.,  2014, \mnras, 438, 3409

\bibitem[\protect\citeauthoryear{{D'Abrusco}, {Fabbiano}, {Djorgovski} \& {et
  al.}}{{D'Abrusco} et~al.}{2012}]{DAbrusco12}
{D'Abrusco} R.,  {Fabbiano} G.,  {Djorgovski} G.,    {et al.} 2012, \apj, 755,
  92

\bibitem[\protect\citeauthoryear{{Dale}, {Gil de Paz}, {Gordon}, {Hanson},
  {Armus}, {Bendo}, {Bianchi}, {Block}, {Boissier} \& et al.}{{Dale}
  et~al.}{2007}]{Dale07}
{Dale} D.~A.,  {Gil de Paz} A.,  {Gordon} K.~D.,  {Hanson} H.~M.,  {Armus} L.,
  {Bendo} G.~J.,  {Bianchi} L.,  {Block} M.,  {Boissier} S.,    et al. 2007,
  \apj, 655, 863

\bibitem[\protect\citeauthoryear{{Delchambre}}{{Delchambre}}{2016}]{Delchambre2016}
{Delchambre} L.,  2016, \mnras, 460, 2811

\bibitem[\protect\citeauthoryear{{Ellison}, {Teimoorinia}, {Rosario} \&
  {Mendel}}{{Ellison} et~al.}{2016}]{Ellison16a}
{Ellison} S.~L.,  {Teimoorinia} H.,  {Rosario} D.~J.,    {Mendel} J.~T.,  2016,
  \mnras, 455, 370

\bibitem[\protect\citeauthoryear{{Folkes}, {Lahav} \& {Maddox}}{{Folkes}
  et~al.}{1996}]{Folkes96}
{Folkes} S.~R.,  {Lahav} O.,    {Maddox} S.~J.,  1996, \mnras, 283, 651

\bibitem[\protect\citeauthoryear{{Fustes}, {Manteiga}, {Dafonte}, {Arcay},
  {Ulla}, {Smith}, {Borrachero} \& {Sordo}}{{Fustes} et~al.}{2013}]{Fustes13}
{Fustes} D.,  {Manteiga} M.,  {Dafonte} C.,  {Arcay} B.,  {Ulla} A.,  {Smith}
  K.,  {Borrachero} R.,    {Sordo} R.,  2013, \aap, 559, A7

\bibitem[\protect\citeauthoryear{{Geach}}{{Geach}}{2012}]{Geach12}
{Geach} J.~E.,  2012, \mnras, 419, 2633

\bibitem[\protect\citeauthoryear{{Gulati}, {Gupta} \& {Singh}}{{Gulati}
  et~al.}{1997}]{Gulati97}
{Gulati} R.,  {Gupta} R.,    {Singh} H.,  1997, \pasp, 109, 843

\bibitem[\protect\citeauthoryear{{Hernandez-Pajares} \&
  {Floris}}{{Hernandez-Pajares} \& {Floris}}{1994}]{Hernandez94}
{Hernandez-Pajares} M.,  {Floris} J.,  1994, \mnras, 268, 444

\bibitem[\protect\citeauthoryear{{Holden}, {Oesch}, {Gonz{\'a}lez},
  {Illingworth}, {Labb{\'e}}, {Bouwens}, {Franx}, {van Dokkum} \&
  {Spitler}}{{Holden} et~al.}{2016}]{Holden16}
{Holden} B.~P.,  {Oesch} P.~A.,  {Gonz{\'a}lez} V.~G.,  {Illingworth} G.~D.,
  {Labb{\'e}} I.,  {Bouwens} R.,  {Franx} M.,  {van Dokkum} P.,    {Spitler}
  L.,  2016, \apj, 820, 73

\bibitem[\protect\citeauthoryear{{Hutchinson}, {Bolton}, {Dawson}, {Allende
  Prieto}, {Bailey}, {Bautista}, {Brownstein}, {Conroy}, {Guy}, {Myers},
  {Newman}, {Prakash}, {Carnero-Rosell}, {Seo}, {Tojeiro}, {Vivek} \& {Ben
  Zhu}}{{Hutchinson} et~al.}{2016}]{Hutchinson2016}
{Hutchinson} T.~A.,  {Bolton} A.~S.,  {Dawson} K.~S.,  {Allende Prieto} C.,
  {Bailey} S.,  {Bautista} J.~E.,  {Brownstein} J.~R.,  {Conroy} C.,  {Guy} J.,
   {Myers} A.~D.,  {Newman} J.~A.,  {Prakash} A.,  {Carnero-Rosell} A.,  {Seo}
  H.-J.,  {Tojeiro} R.,  {Vivek} M.,    {Ben Zhu} G.,  2016, \aj, 152, 205

\bibitem[\protect\citeauthoryear{{Ilbert} et~al.,}{{Ilbert}
  et~al.}{2006}]{Ilbert2006}
{Ilbert} O.,  et~al., 2006, \aap, 457, 841

\bibitem[\protect\citeauthoryear{{in der Au}, {Meusinger}, {Schalldach} \&
  {Newholm}}{{in der Au} et~al.}{2012}]{In12}
{in der Au} A.,  {Meusinger} H.,  {Schalldach} P.~F.,    {Newholm} M.,  2012,
  \aap, 547, A115

\bibitem[\protect\citeauthoryear{{Karampelas}, {Kontizas}, {Rocca-Volmerange},
  {Bellas-Velidis}, {Kontizas}, {Livanou}, {Tsalmantza} \&
  {Dapergolas}}{{Karampelas} et~al.}{2012}]{Karampelas2012}
{Karampelas} A.,  {Kontizas} M.,  {Rocca-Volmerange} B.,  {Bellas-Velidis} I.,
  {Kontizas} E.,  {Livanou} E.,  {Tsalmantza} P.,    {Dapergolas} A.,  2012,
  \aap, 538, A38

\bibitem[\protect\citeauthoryear{{Kinney}, {Bohlin}, {Calzetti}, {Panagia} \&
  {Wyse}}{{Kinney} et~al.}{1993}]{Kinney93}
{Kinney} A.~L.,  {Bohlin} R.~C.,  {Calzetti} D.,  {Panagia} N.,    {Wyse}
  R.~F.~G.,  1993, \apjs, 86, 5

\bibitem[\protect\citeauthoryear{{Kinney}, {Calzetti}, {Bohlin} \& {et
  al.}}{{Kinney} et~al.}{1996}]{Kinney96}
{Kinney} A.~L.,  {Calzetti} D.,  {Bohlin} R.~C.,    {et al.} 1996, \apj, 467,
  38

\bibitem[\protect\citeauthoryear{Kohonen}{Kohonen}{1982}]{Kohonen82}
Kohonen T.,  1982, Biological Cybernetics, 43, 59

\bibitem[\protect\citeauthoryear{Kohonen}{Kohonen}{1998}]{Kohonen98}
Kohonen T.,  1998, Neurocomputing, 21, 1

\bibitem[\protect\citeauthoryear{{Laporte}, {Infante}, {Troncoso Iribarren},
  {Zheng}, {Molino} \& {et al.}}{{Laporte} et~al.}{2016}]{Laporte16}
{Laporte} N.,  {Infante} L.,  {Troncoso Iribarren} P.,  {Zheng} W.,  {Molino}
  A.,    {et al.} 2016, \apj, 820, 98

\bibitem[\protect\citeauthoryear{MacQueen}{MacQueen}{1967}]{Macqueen67}
MacQueen J.,  1967, in Proceedings of the Fifth Berkeley Symposium on
  Mathematical Statistics and Probability, Volume 1: Statistics Some methods
  for classification and analysis of multivariate observations.
University of California Press, Berkeley, Calif., pp 281--297

\bibitem[\protect\citeauthoryear{{Maehoenen} \& {Hakala}}{{Maehoenen} \&
  {Hakala}}{1995}]{Maehoenen95}
{Maehoenen} P.~H.,  {Hakala} P.~J.,  1995, \apjl, 452, L77

\bibitem[\protect\citeauthoryear{{Mannucci}, {Basile}, {Poggianti}, {Cimatti},
  {Daddi}, {Pozzetti} \& {Vanzi}}{{Mannucci} et~al.}{2001}]{Mannucci01}
{Mannucci} F.,  {Basile} F.,  {Poggianti} B.~M.,  {Cimatti} A.,  {Daddi} E.,
  {Pozzetti} L.,    {Vanzi} L.,  2001, \mnras, 326, 745

\bibitem[\protect\citeauthoryear{{Marquez}, {Hill}, {Worthley} \&
  {Remus}}{{Marquez} et~al.}{1991}]{Marquez91}
{Marquez} L.,  {Hill} T.,  {Worthley} R.,    {Remus} W.,  1991, in Proceedings
  of the Twenty-Fourth Annual Hawaii International Conference on System
  Sciences Vol.~4, Neural network models as an alternative to regression.
pp 129--135

\bibitem[\protect\citeauthoryear{{Meusinger}, {Schalldach}, {Mirhosseini} \&
  {Pertermann}}{{Meusinger} et~al.}{2016}]{Meusinger16}
{Meusinger} H.,  {Schalldach} P.,  {Mirhosseini} A.,    {Pertermann} F.,  2016,
  \aap, 587, A83

\bibitem[\protect\citeauthoryear{{Miller} \& {Coe}}{{Miller} \&
  {Coe}}{1996}]{Miller96}
{Miller} A.~S.,  {Coe} M.~J.,  1996, \mnras, 279, 293

\bibitem[\protect\citeauthoryear{{Murtagh}}{{Murtagh}}{1995}]{Murtagh95}
{Murtagh} F.,  1995, in {Shaw} R.~A.,  {Payne} H.~E.,   {Hayes} J.~J.~E.,  eds,
  Astronomical Data Analysis Software and Systems IV Vol.~77 of Astronomical
  Society of the Pacific Conference Series, {Unsupervised Catalog
  Classification}.
p.~264

\bibitem[\protect\citeauthoryear{{Noll}, {Burgarella}, {Giovannoli}, {Buat},
  {Marcillac} \& {Mu{\~n}oz-Mateos}}{{Noll} et~al.}{2009}]{Noll09}
{Noll} S.,  {Burgarella} D.,  {Giovannoli} E.,  {Buat} V.,  {Marcillac} D.,
  {Mu{\~n}oz-Mateos} J.~C.,  2009, \aap, 507, 1793

\bibitem[\protect\citeauthoryear{{Odewahn}, {Stockwell}, {Pennington},
  {Humphreys} \& {Zumach}}{{Odewahn} et~al.}{1992}]{Odewahn92}
{Odewahn} S.~C.,  {Stockwell} E.~B.,  {Pennington} R.~L.,  {Humphreys} R.~M.,
   {Zumach} W.~A.,  1992, \aj, 103, 318

\bibitem[\protect\citeauthoryear{{Ordov{\'a}s-Pascual} \& {S{\'a}nchez
  Almeida}}{{Ordov{\'a}s-Pascual} \& {S{\'a}nchez Almeida}}{2014}]{Ordov14}
{Ordov{\'a}s-Pascual} I.,  {S{\'a}nchez Almeida} J.,  2014, \aaj, 565, A53

\bibitem[\protect\citeauthoryear{{Paiano}, {Landoni}, {Falomo}, {Scarpa} \&
  {Treves}}{{Paiano} et~al.}{2016}]{Paiano16}
{Paiano} S.,  {Landoni} M.,  {Falomo} R.,  {Scarpa} R.,    {Treves} A.,  2016,
  \mnras, 458, 2836

\bibitem[\protect\citeauthoryear{Pedregosa, Varoquaux, Gramfort
  et~al.,}{Pedregosa et~al.}{2011}]{sklearn}
Pedregosa F.,  Varoquaux G.,  Gramfort A.,    et~al., 2011, Journal of Machine
  Learning Research, 12, 2825

\bibitem[\protect\citeauthoryear{{Rajaniemi} \& {M{\"a}h{\"o}nen}}{{Rajaniemi}
  \& {M{\"a}h{\"o}nen}}{2002}]{Rajaniemi02}
{Rajaniemi} H.~J.,  {M{\"a}h{\"o}nen} P.,  2002, \apj, 566, 202

\bibitem[\protect\citeauthoryear{Rousseeuw}{Rousseeuw}{1987}]{rousseeuw87}
Rousseeuw P.~J.,  1987, Journal of Computational and Applied Mathematics, 20,
  53

\bibitem[\protect\citeauthoryear{{Salpeter}}{{Salpeter}}{1955}]{Salpeter55}
{Salpeter} E.~E.,  1955, \apj, 121, 161

\bibitem[\protect\citeauthoryear{{Santini}, {Fontana}, {Grazian}, {Salimbeni},
  {Fiore}, {Fontanot}, {Boutsia}, {Castellano}, {Cristiani}, {de Santis},
  {Gallozzi}, {Giallongo}, {Menci}, {Nonino}, {Paris}, {Pentericci} \&
  {Vanzella}}{{Santini} et~al.}{2009}]{Santini09}
{Santini} P.,  {Fontana} A.,  {Grazian} A.,  {Salimbeni} S.,  {Fiore} F.,
  {Fontanot} F.,  {Boutsia} K.,  {Castellano} M.,  {Cristiani} S.,  {de Santis}
  C.,  {Gallozzi} S.,  {Giallongo} E.,  {Menci} N.,  {Nonino} M.,  {Paris} D.,
  {Pentericci} L.,    {Vanzella} E.,  2009, \aap, 504, 751

\bibitem[\protect\citeauthoryear{{Scaringi}, {Cottis}, {Knigge} \&
  {Goad}}{{Scaringi} et~al.}{2009}]{Scaringi09}
{Scaringi} S.,  {Cottis} C.~E.,  {Knigge} C.,    {Goad} M.~R.,  2009, \mnras,
  399, 2231

\bibitem[\protect\citeauthoryear{Seber}{Seber}{1984}]{Seber84}
Seber G., , 1984, Multivariate analysis

\bibitem[\protect\citeauthoryear{{Shakouri}, {Johnston-Hollitt} \&
  {Dehghan}}{{Shakouri} et~al.}{2016}]{Shakouri16}
{Shakouri} S.,  {Johnston-Hollitt} M.,    {Dehghan} S.,  2016, \mnras, 458,
  3083

\bibitem[\protect\citeauthoryear{{Shi}, {Liu}, {Sun}, {Li}, {Lei} \&
  {Wang}}{{Shi} et~al.}{2015}]{Shi15}
{Shi} F.,  {Liu} Y.-Y.,  {Sun} G.-L.,  {Li} P.-Y.,  {Lei} Y.-M.,    {Wang} J.,
  2015, \mnras, 453, 122

\bibitem[\protect\citeauthoryear{Spath}{Spath}{1985}]{Spath85}
Spath H.,  1985, The cluster dissection and analysis theory FORTRAN programs
  examples.
Prentice-Hall, Inc.

\bibitem[\protect\citeauthoryear{{Tammour}, {Gallagher}, {Daley} \&
  {Richards}}{{Tammour} et~al.}{2016}]{Aycha16}
{Tammour} A.,  {Gallagher} S.~C.,  {Daley} M.,    {Richards} G.~T.,  2016,
  \mnras, 459, 1659

\bibitem[\protect\citeauthoryear{{Teimoorinia}}{{Teimoorinia}}{2012}]{Hossein12}
{Teimoorinia} H.,  2012, \aj, 144, 172

\bibitem[\protect\citeauthoryear{{Teimoorinia}, {Bluck} \&
  {Ellison}}{{Teimoorinia} et~al.}{2016}]{Hossein16a}
{Teimoorinia} H.,  {Bluck} A.~F.~L.,    {Ellison} S.~L.,  2016, \mnras, 457,
  2086

\bibitem[\protect\citeauthoryear{{Vanzella}, {Cristiani}, {Dickinson},
  {Giavalisco}, {Kuntschner}, {Haase}, {Nonino} \& {GOODS Team}}{{Vanzella}
  et~al.}{2008}]{Vanzella08}
{Vanzella} E.,  {Cristiani} S.,  {Dickinson} M.,  {Giavalisco} M.,
  {Kuntschner} H.,  {Haase} J.,  {Nonino} M.,    {GOODS Team} 2008, \aap, 478,
  83

\bibitem[\protect\citeauthoryear{{Vanzella}, {Cristiani}, {Dickinson} \& {GOODS
  Team}}{{Vanzella} et~al.}{2005}]{Vanzella05}
{Vanzella} E.,  {Cristiani} S.,  {Dickinson} M.,    {GOODS Team} 2005, \aap,
  434, 53

\bibitem[\protect\citeauthoryear{{Vanzella}, {Cristiani}, {Dickinson} \& {GOODS
  Team}}{{Vanzella} et~al.}{2006}]{Vanzella06}
{Vanzella} E.,  {Cristiani} S.,  {Dickinson} M.,    {GOODS Team} 2006, \aap,
  454, 423

\bibitem[\protect\citeauthoryear{{Vanzella}, {Cristiani}, {Fontana}, {Nonino},
  {Arnouts}, {Giallongo}, {Grazian}, {Fasano}, {Popesso}, {Saracco} \&
  {Zaggia}}{{Vanzella} et~al.}{2004}]{vanzella04}
{Vanzella} E.,  {Cristiani} S.,  {Fontana} A.,  {Nonino} M.,  {Arnouts} S.,
  {Giallongo} E.,  {Grazian} A.,  {Fasano} G.,  {Popesso} P.,  {Saracco} P.,
  {Zaggia} S.,  2004, \aap, 423, 761

\bibitem[\protect\citeauthoryear{{Vesanto}}{{Vesanto}}{2005}]{Vesanto05}
{Vesanto} J., , 2005, {SOM implementation in SOM Toolbox},
  \url{http://www.cis.hut.fi/projects/somtoolbox/documentation/somalg.shtml}(accessed
  MARCH 2, 2016)

\end{thebibliography}

\newpage
\appendix
\section{Inputs to galaxy spectral classification}
\label{app:sed_props}

The spectra classified in this work are available through the Supplementary Information to this paper.
The data are contained in a single machine-readable file with wavelength as the rows and individual galaxy spectra as the columns.
Table~\ref{tab:sed} gives the properties of the best-fit models used to create these spectra.

\begin{table*}
\caption{Properties of best-fit models created by T12.
The full table is available online.}
\label{tab:sed}
\centering
\begin{tabular}{llcrrrrcc}
\hline\hline
\noalign{\smallskip}
ID & GOODS ID & redshift & $\log(M_\mathrm{star})$ [M$_\odot$] & $t_{\,\mathrm{D4000}}$ [Gyr] & $\log(\phi)$ [Gyr$^{-1}$] &$\log(f_\mathrm{burst})$& $A_\mathrm{FUV}$ & T12 class\\
\noalign{\smallskip}
\hline
\noalign{\smallskip}
G1 & GDS-J033214.59-274913.4 & 0.468 & 9.09 $\pm$ 0.29 & -1.42 $\pm$ 0.1 & 0.30 $\pm$ 0.32 & -0.32 $\pm$ 0.45 & 0.70 $\pm$ 0.45 & SB-1 \\
G2 & GDS-J033227.88-275140.4 & 0.521 & 11.11 $\pm$ 0.06 & 0.66 $\pm$ 0.05 & -2.36 $\pm$ 0.19 & -3.19 $\pm$ 0.41 & 0.83 $\pm$ 0.29 &  E \\
G3 & GDS-J033244.7-274922.8 & 0.522 & 10.56 $\pm$0.05 & -0.49 $\pm$ 0.01 & -1.26 $\pm$ 0.07 & -1.00 $\pm$ 0.04 & 1.56 $\pm$ 0.09 &  --- \\
G4 & GDS-J033213.21-274715.7 & 0.523 & 8.78 $\pm$ 0.31 & -1.18 $\pm$ 0.34 & -0.05 $\pm$ 0.49 & -0.70 $\pm$ 0.73 & 2.26 $\pm$ 0.95 &  --- \\
G5 & GDS-J033230.23-274519.9 & 0.523 & 9.12 $\pm$ 0.28 & -1.27 $\pm$ 0.2 & 0.07 $\pm$0.33 & -0.35 $\pm$ 0.46 & 1.34 $\pm$ 0.57 & SB-4 \\
G6 & GDS-J033232.18-274534.9 & 0.523 & 11.23 $\pm$ 0.04 & 0.41 $\pm$ 0.04 & -1.70 $\pm$ 0.09 & -3.00 $\pm$0.09 & 0.88 $\pm$ 0.04 &  --- \\
\noalign{\smallskip}
\hline
\end{tabular}
\end{table*}

\newpage
\section{Results of training for one-dimensional self-organizing maps}
\label{app: high_Z_1d_soms}
As mentioned in Sec.~\ref{sec: 1D_somz}, we changed the size of the SOM from $1\times2$ to $1\times22$. In that section we show some example of the results; here we show the rest of the maps to monitor the changes in the map over various sizes.

\label{app: 1d}
    \begin{figure}
        \begin{subfigure}[b]{0.5\textwidth}
            \centering
            \includegraphics[width=\textwidth]{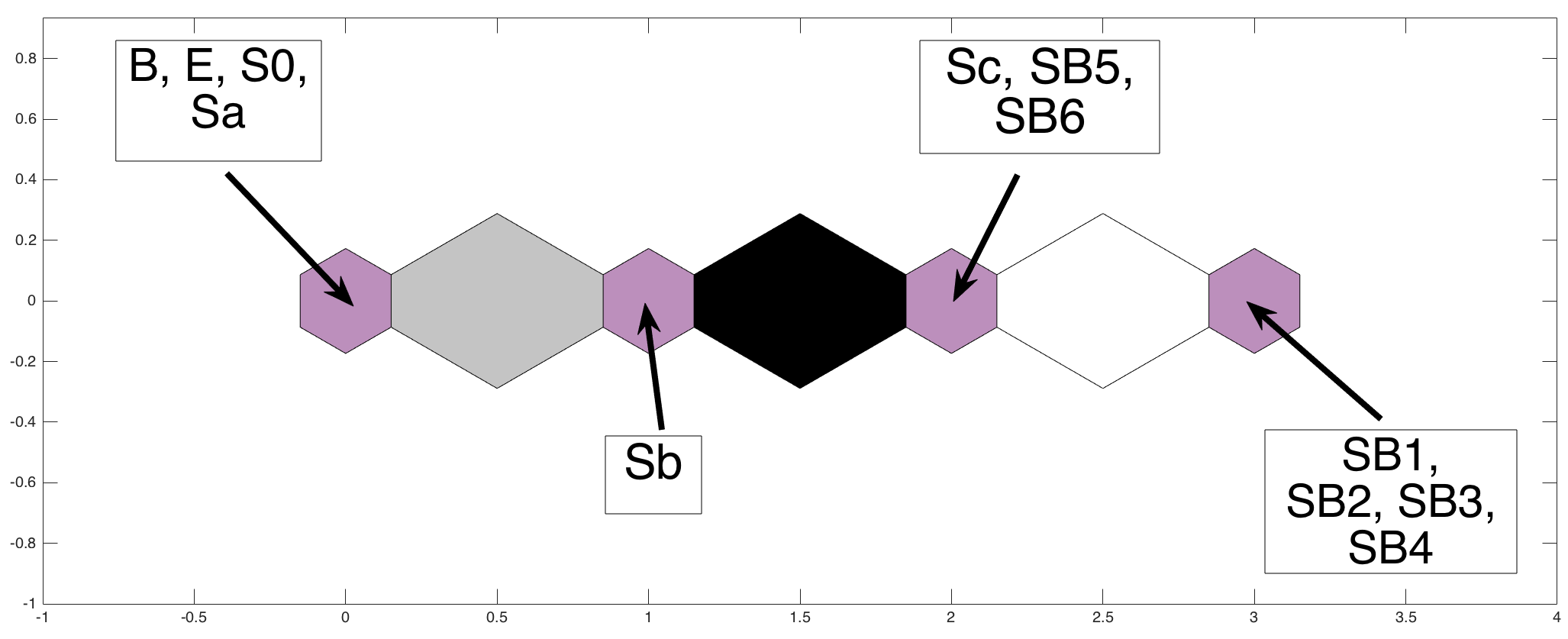}
        \end{subfigure}
        \hfill
        \begin{subfigure}[b]{0.5\textwidth}
             \includegraphics[width=\textwidth]{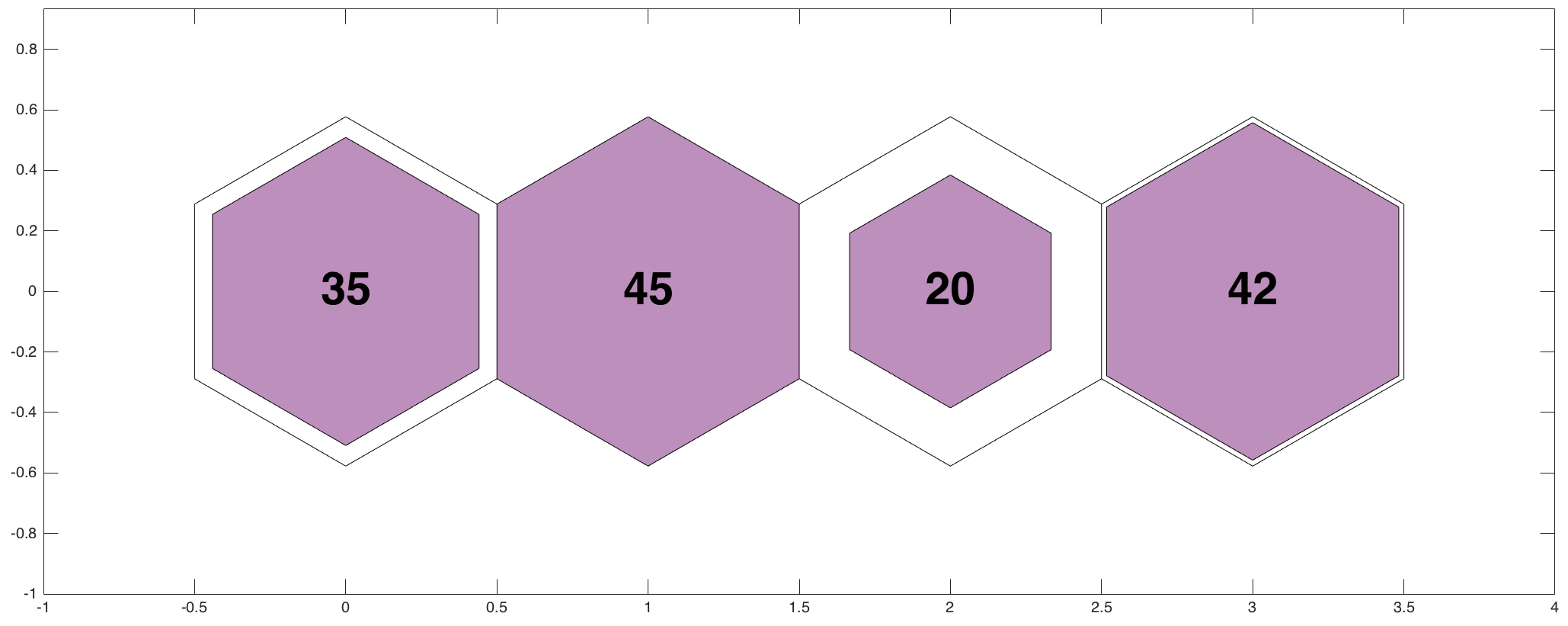}
        \end{subfigure}
                \caption{Results of training network in $1\times4$~grid.}
         \label{fig: 1by4T}
    \end{figure}

    \begin{figure}
        \begin{subfigure}[b]{0.5\textwidth}
            \centering
            \includegraphics[width=\textwidth]{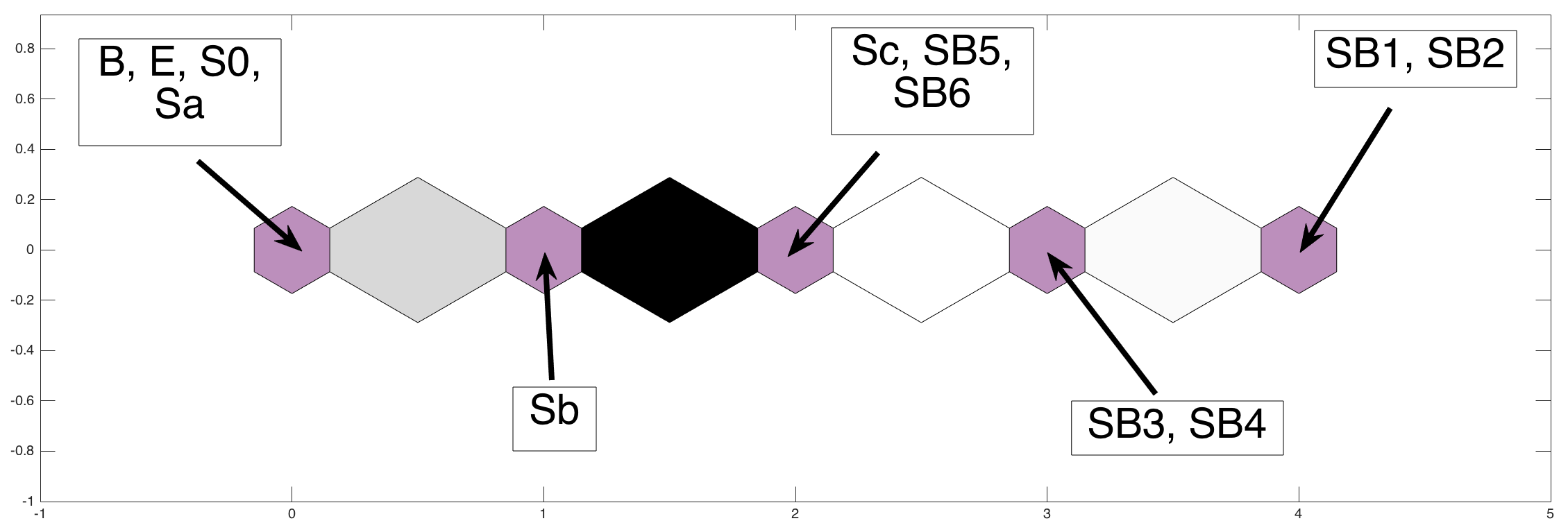}
        \end{subfigure}
        \hfill
        \begin{subfigure}[b]{0.5\textwidth}
             \includegraphics[width=\textwidth]{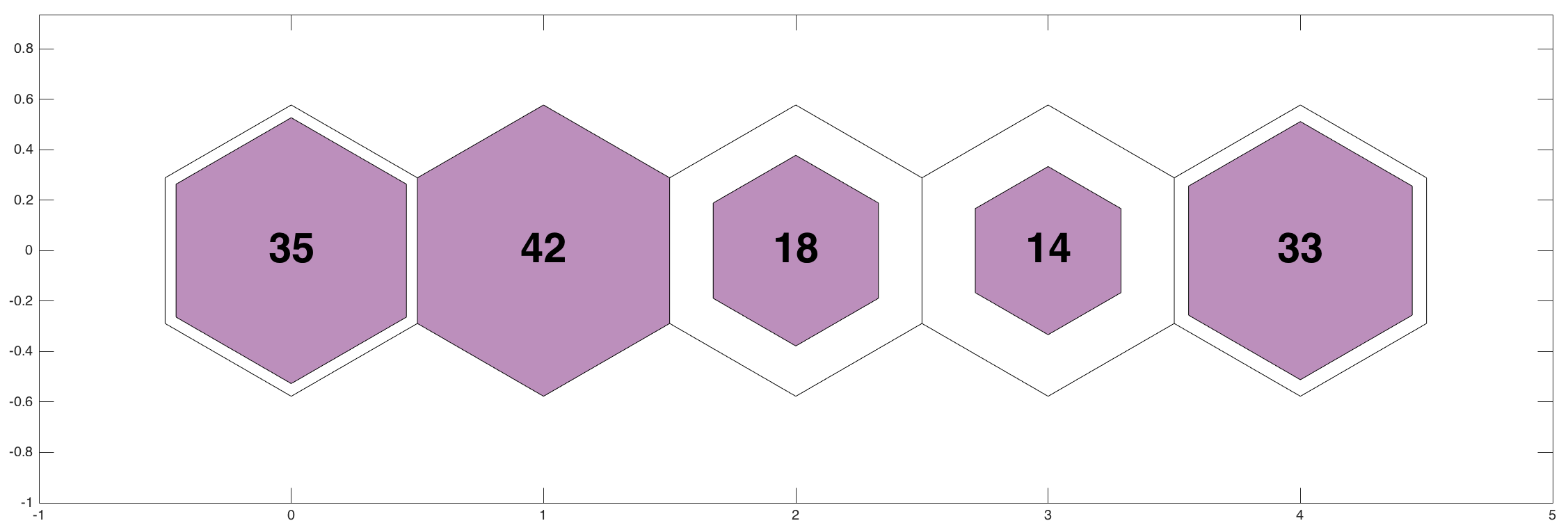}
        \end{subfigure}
                \caption{Results of training network in $1\times5$~grid.}
         \label{fig: 1by5T}
    \end{figure}

    \begin{figure}
        \begin{subfigure}[b]{0.5\textwidth}
            \centering
            \includegraphics[width=\textwidth]{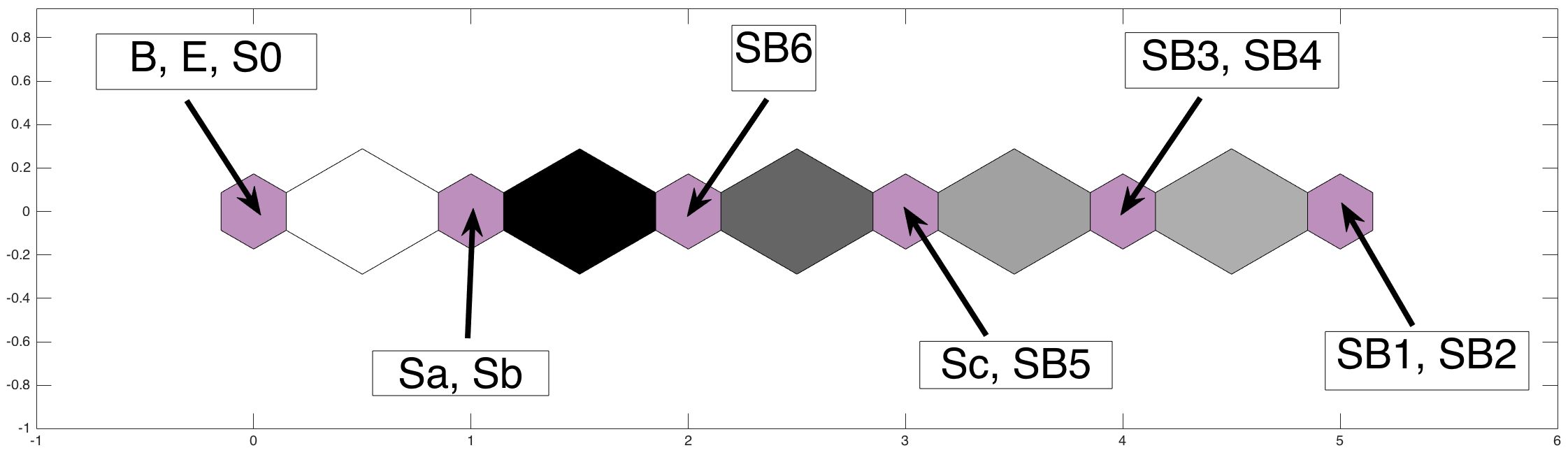}
        \end{subfigure}
        \hfill
        \begin{subfigure}[b]{0.5\textwidth}
             \includegraphics[width=\textwidth]{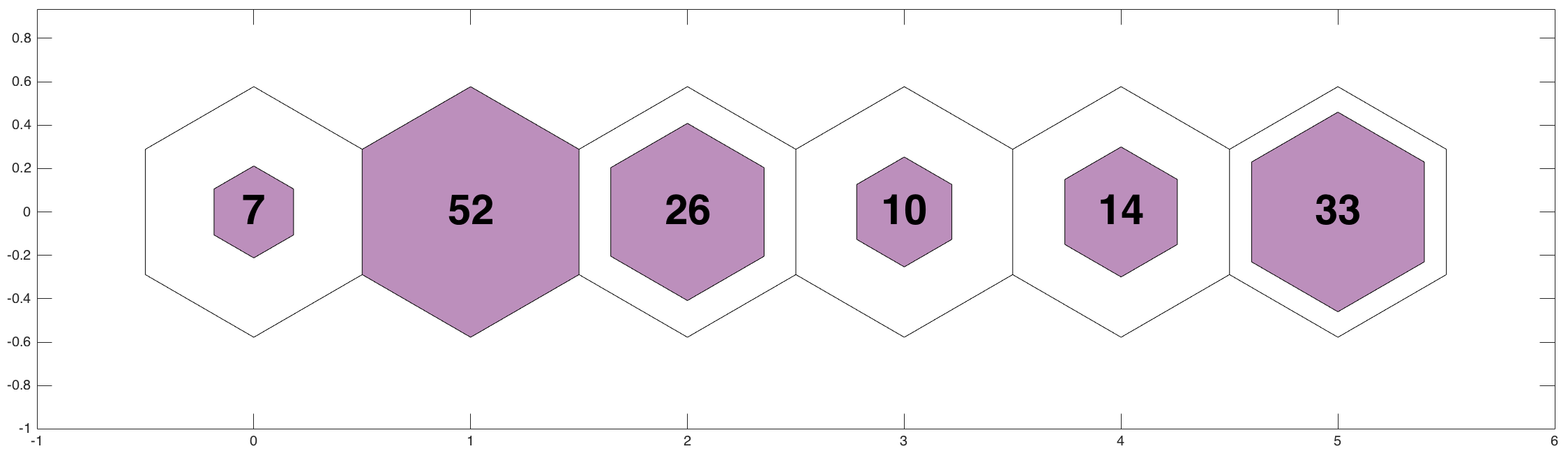}
        \end{subfigure}
                \caption{Results of training network in $1\times6$~grid.}
         \label{fig: 1by6T}
    \end{figure}

    \begin{figure}
        \begin{subfigure}[b]{0.5\textwidth}
            \centering
            \includegraphics[width=\textwidth]{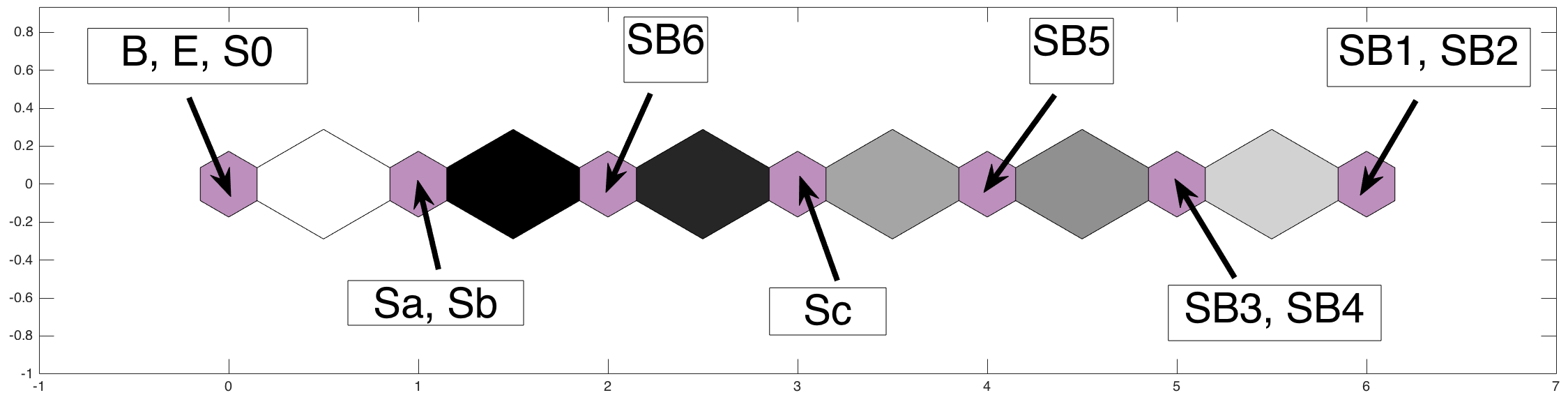}
        \end{subfigure}
        \hfill
        \begin{subfigure}[b]{0.5\textwidth}
             \includegraphics[width=\textwidth]{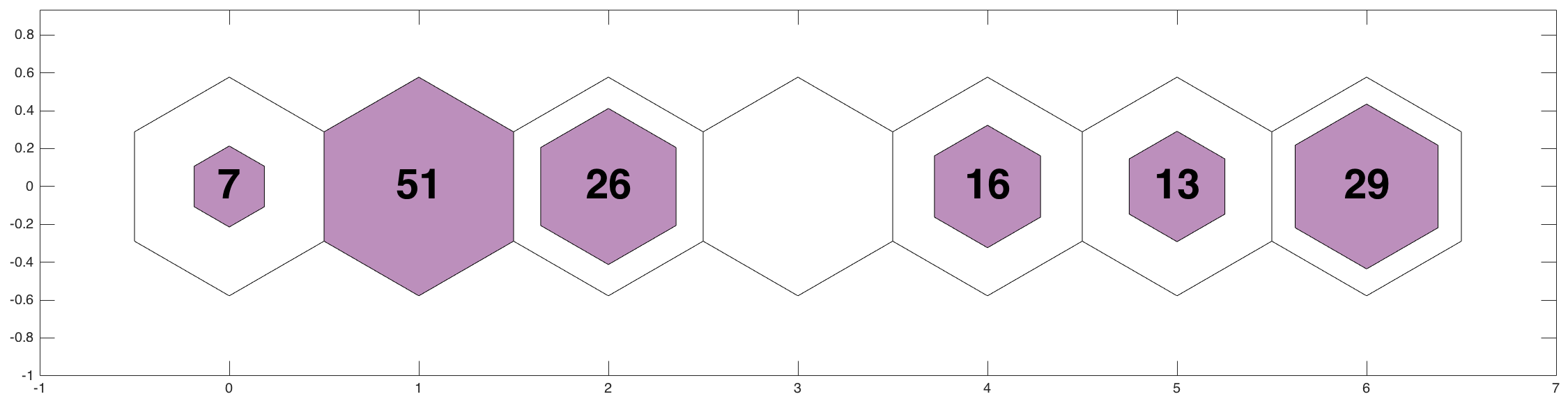}
        \end{subfigure}
                \caption{Results of training network in $1\times7$~grid.}
         \label{fig: 1by7T}
    \end{figure}

    \begin{figure}
        \begin{subfigure}[b]{0.5\textwidth}
            \centering
            \includegraphics[width=\textwidth]{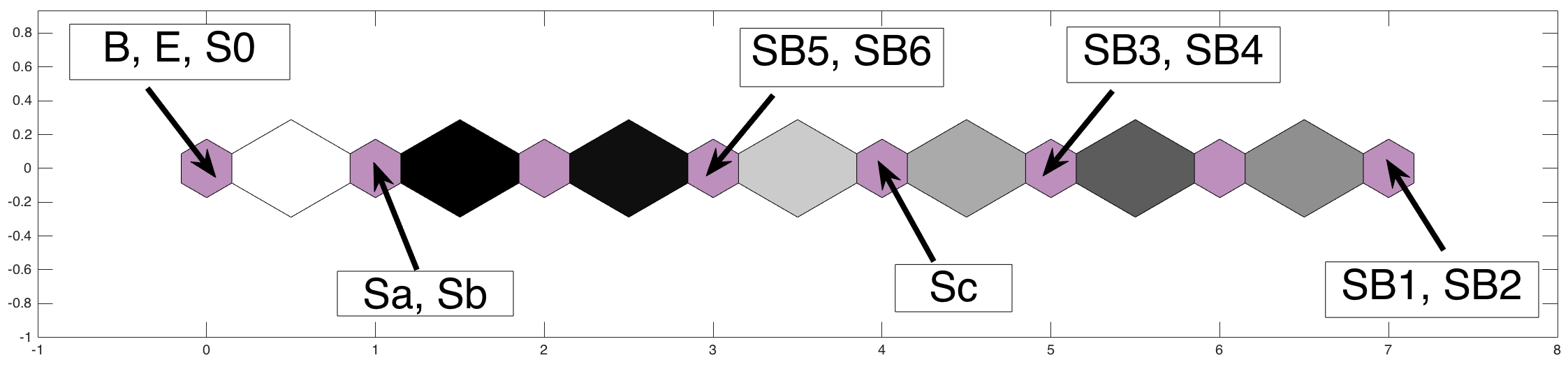}
        \end{subfigure}
        \hfill
        \begin{subfigure}[b]{0.5\textwidth}
             \includegraphics[width=\textwidth]{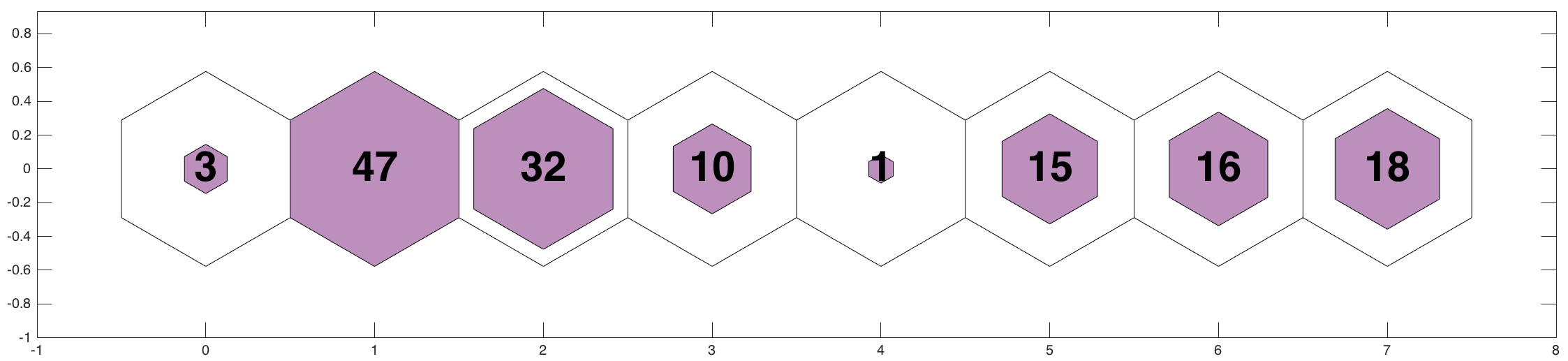}
        \end{subfigure}
                \caption{Results of training network in $1\times8$~grid.}
         \label{fig: 1by8T}
    \end{figure}

    \begin{figure}
        \begin{subfigure}[b]{0.5\textwidth}
            \centering
            \includegraphics[width=\textwidth]{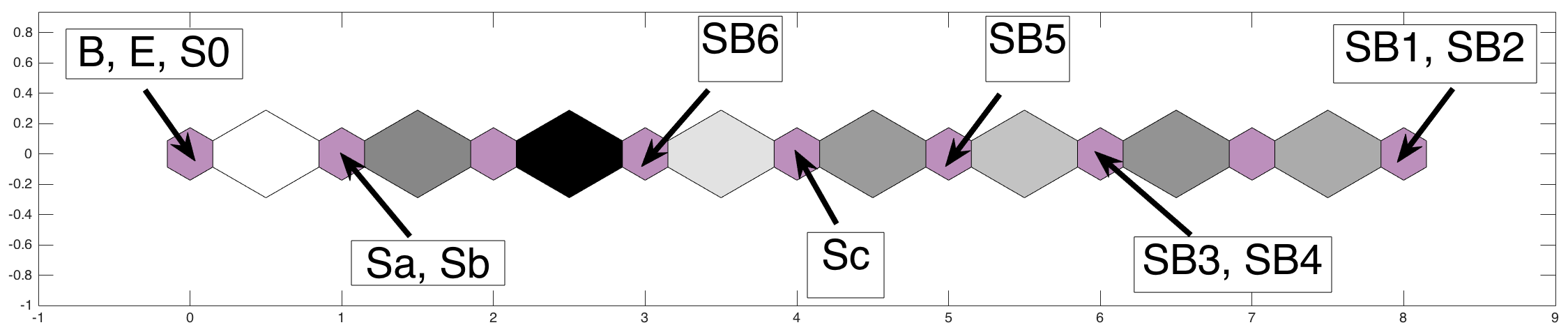}
        \end{subfigure}
        \hfill
        \begin{subfigure}[b]{0.5\textwidth}
             \includegraphics[width=\textwidth]{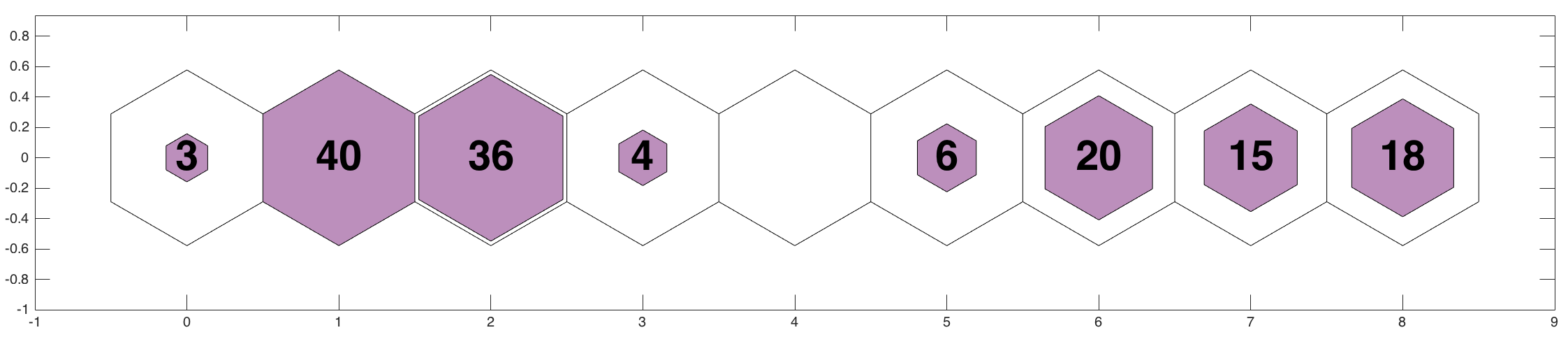}
        \end{subfigure}
                \caption{Results of training network in $1\times9$~grid.}
         \label{fig: 1by9T}
    \end{figure}

    \begin{figure}
        \begin{subfigure}[b]{0.5\textwidth}
            \centering
            \includegraphics[width=\textwidth,height=2.5cm]{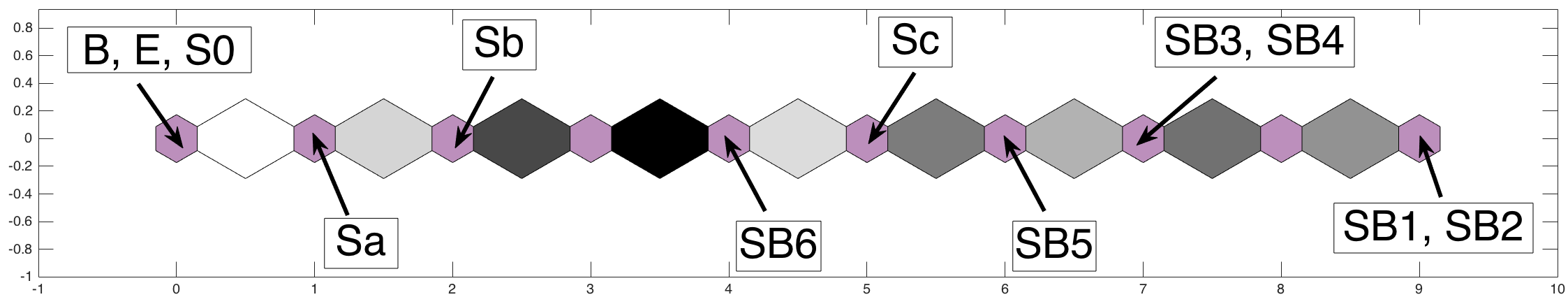}
        \end{subfigure}
        \hfill
        \begin{subfigure}[b]{0.5\textwidth}
             \includegraphics[width=\textwidth,height=2.5cm]{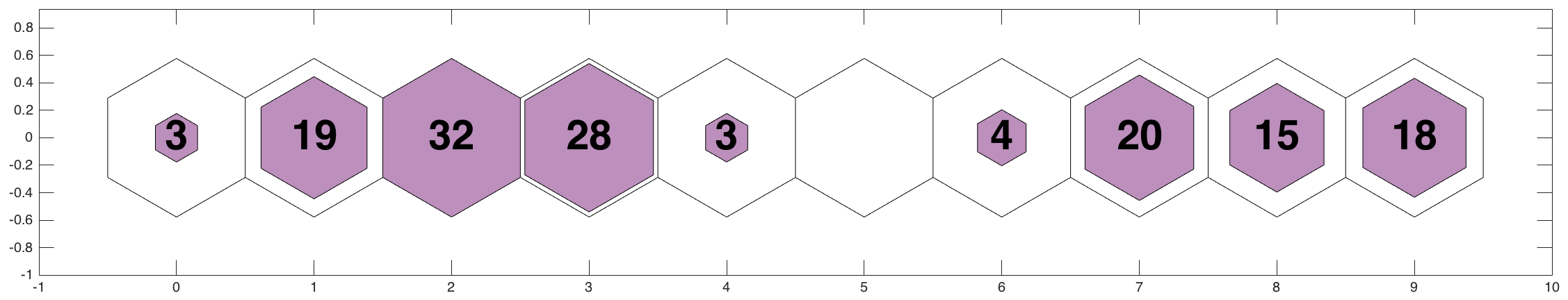}
        \end{subfigure}
                \caption{Results of training network in $1\times10$~grid.}
         \label{fig: 1by10T}
    \end{figure}

    \begin{figure}
        \begin{subfigure}[b]{0.5\textwidth}
            \centering
            \includegraphics[width=\textwidth,height=2.5cm]{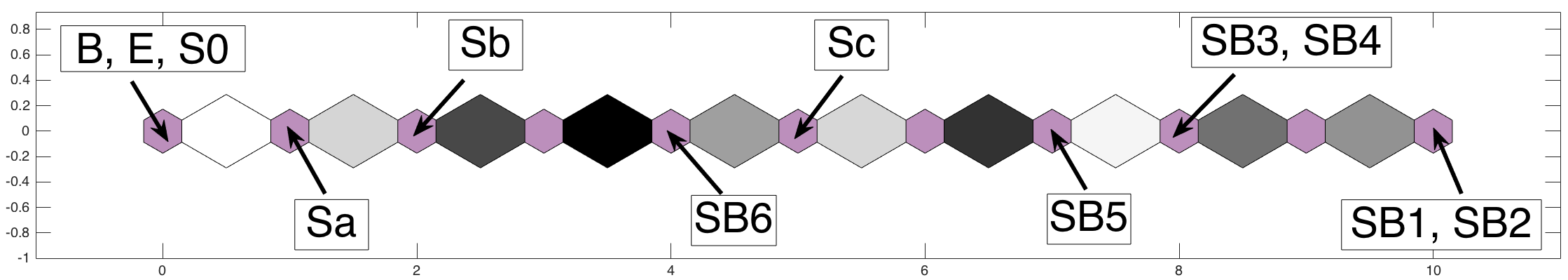}
        \end{subfigure}
        \hfill
        \begin{subfigure}[b]{0.5\textwidth}
             \includegraphics[width=\textwidth,height=2.5cm]{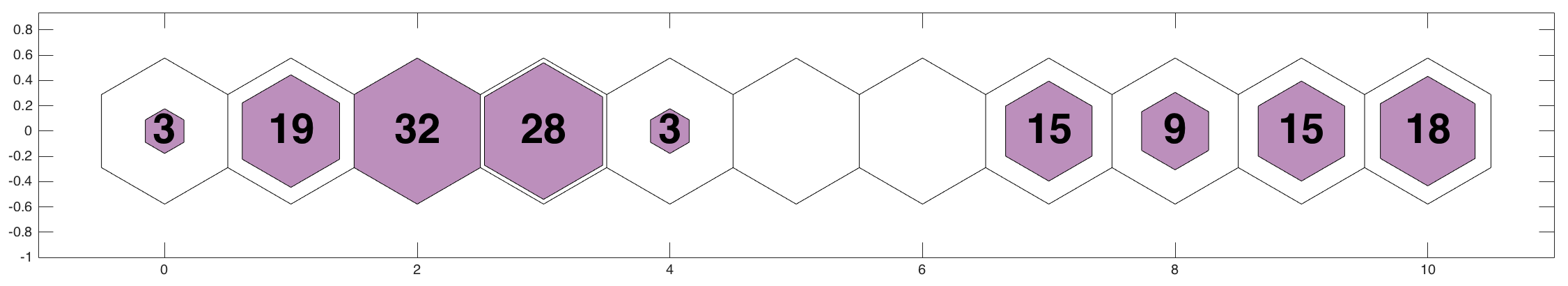}
        \end{subfigure}
                \caption{Results of training network in $1\times11$~grid.}
         \label{fig: 1by11T}
    \end{figure}

    \begin{figure}
        \begin{subfigure}[b]{0.5\textwidth}
            \centering
            \includegraphics[width=\textwidth,height=2.5cm]{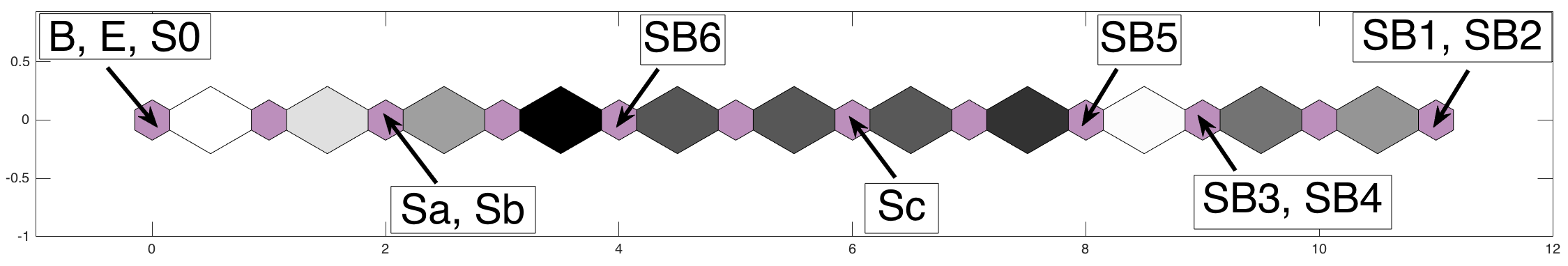}
        \end{subfigure}
        \hfill
        \begin{subfigure}[b]{0.5\textwidth}
             \includegraphics[width=\textwidth,height=2.5cm]{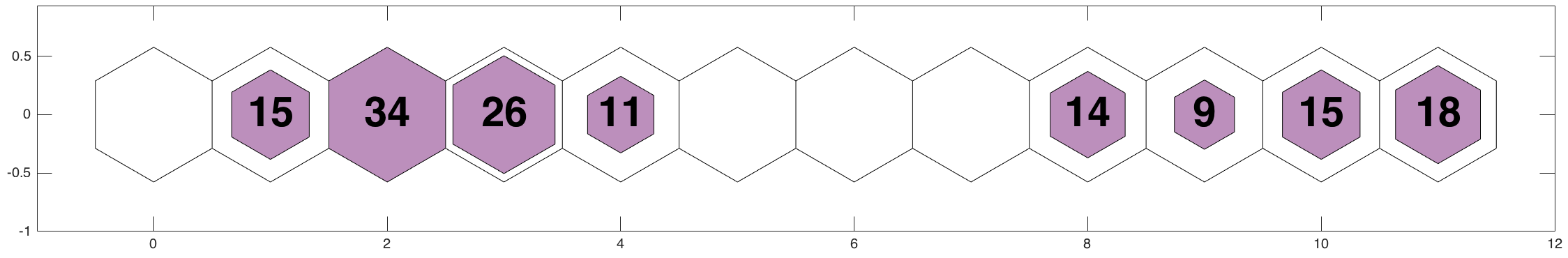}
        \end{subfigure}
                \caption{Results of training network in $1\times12$~grid.}
         \label{fig: 1by12T}
    \end{figure}

    \begin{figure}
        \begin{subfigure}[b]{0.5\textwidth}
            \centering
            \includegraphics[width=\textwidth,height=2.5cm]{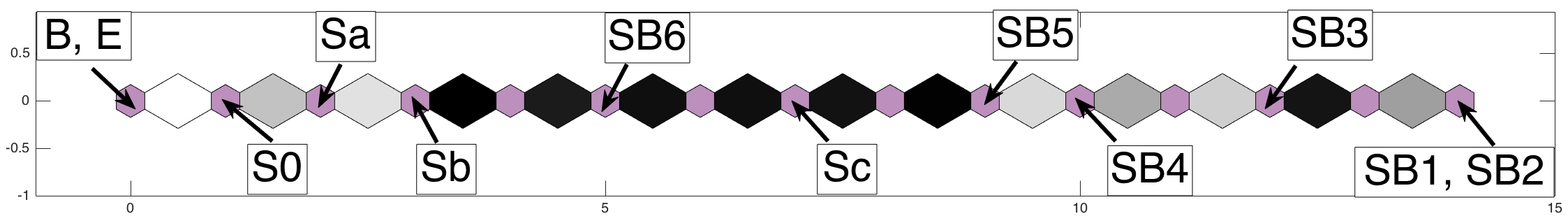}
        \end{subfigure}
        \hfill
        \begin{subfigure}[b]{0.5\textwidth}
             \includegraphics[width=\textwidth,height=2.5cm]{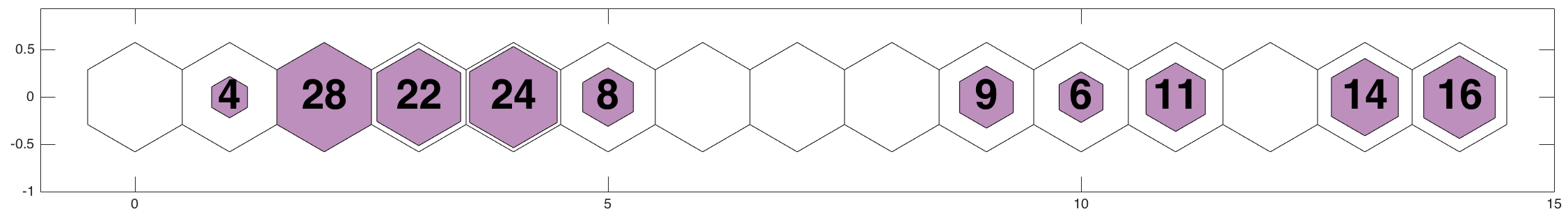}
        \end{subfigure}
                \caption{Results of training network in $1\times15$~grid.}
         \label{fig: 1by15T}
    \end{figure}

    \begin{figure}
        \begin{subfigure}[b]{0.5\textwidth}
            \centering
            \includegraphics[width=\textwidth,height=2.5cm]{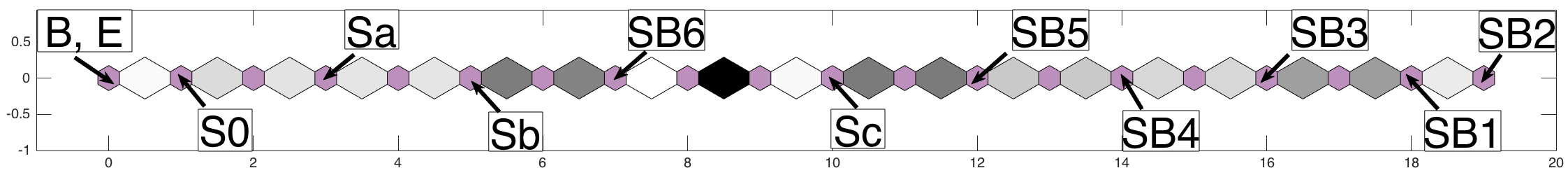}
        \end{subfigure}
        \hfill
        \begin{subfigure}[b]{0.5\textwidth}
             \includegraphics[width=\textwidth,height=2.5cm]{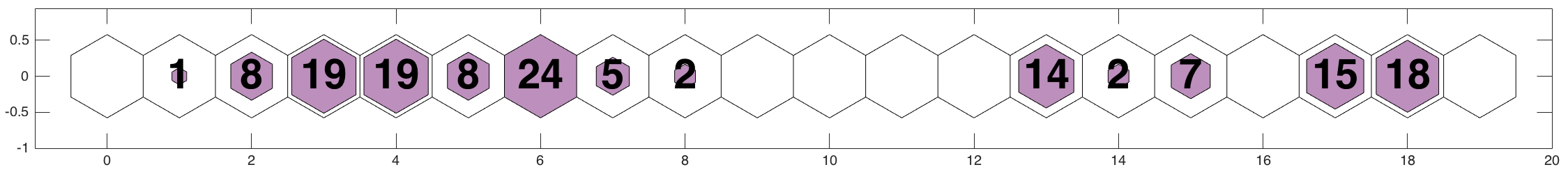}
        \end{subfigure}
                \caption{Results of training network in $1\times20$~grid.}
         \label{fig: 1by20T}
    \end{figure}

\newpage

\end{document}